\documentclass[12 pt, amsfonts, amssymb,amsthm,color]{article} 
\usepackage{mathtools}

\usepackage{setspace}
\usepackage[colorlinks]{hyperref} 
\usepackage[all]{xy}
\usepackage{graphics,graphicx,amsbsy,amssymb,color}
\usepackage[top=2cm, bottom=3cm, left=2cm, right=2cm]{geometry}

\title{\textbf{
On Geometry of Dissipation in Multiscale Dynamics and Thermodynamics}}

 \date{}

\begin{document}

\maketitle

\begin{center} 
 Begüm Ateşli$^{\ast,\ast\ast}$\footnote{e-mail: 
\href{mailto:b.atesli@gtu.edu.tr}{b.atesli@gtu.edu.tr}, \href{mailto:begumatesli@itu.edu.tr}{begumatesli@itu.edu.tr} corresponding author,}, 
O\u{g}ul Esen$^{\ast\ast,\dagger}$\footnote{e-mail: \href{mailto:oesen@gtu.edu.tr}{oesen@gtu.edu.tr},}, 
Miroslav Grmela$^{\dagger\dagger}$\footnote{e-mail: \href{mailto:miroslav.grmela@polymtl.ca} {miroslav.grmela@polymtl.ca},}, 
  Michal Pavelka$^{\ddagger}$\footnote{e-mail: \href{mailto:pavelka@karlin.mff.cuni.cz}{pavelka@karlin.mff.cuni.cz}} 
  \\

\bigskip

$^{\ast}$Department of Mathematics Engineering \\ İstanbul Technical University,  34467 Maslak, İstanbul, Turkey

\bigskip

$^{\ast\ast}$Department of Mathematics, \\ Gebze Technical University, 41400 Gebze,
Kocaeli, Turkey
\bigskip

$^{\dagger}$Center for Mathematics and its Applications,  \\ 
Khazar University, Baku, AZ1096, Azerbaijan
\bigskip

$^{\dagger\dagger}$\'{E}cole Polytechnique de Montr\'{e}al,
  C.P.6079 suc. Centre-ville, 
 \\  Montr\'{e}al, H3C 3A7,  Qu\'{e}bec, Canada
\bigskip

$^{\ddagger}$Mathematical Institute, Faculty of Mathematics and Physics,  \\  Charles University, 
 Sokolovsk\'{a} 83, 18675 Prague, Czech Republic

\begin{abstract}
This manuscript introduces novel approaches to three phenomena. First, we extend the algebraic formulation of kinetic theory within the contact framework by making explicit the gauge freedom, thereby obtaining a formulation in which the phase-space volume itself becomes an additional dynamical variable. Second, we develop a new and simpler geometric formulation of the GENERIC framework, unifying Hamiltonian and gradient dynamics in a contact-geometric setting. This is realized within a specifically constructed graph space, which naturally emerges as an intermediate structure in the geometric Hamilton–Jacobi framework. Finally, we formulate a geometric extension of non-equilibrium thermodynamics in the setting of geometric Hamilton–Jacobi theory, allowing for the inclusion of microturbulence — a key feature of complex dynamical systems.


\smallskip 

\smallskip


\end{abstract}

\end{center}

\tableofcontents
\setlength{\parskip}{4mm}

\onehalfspacing

\section{Introduction}

The use of contact geometry in understanding the geometric foundations of thermodynamics—from its early formulations in the 1970s to its current developments—has been explored in numerous studies, including the (necessarily incomplete) list \cite{bravetti19,goshtermo,lychagin2019contact,chen1998second,mrugala,mrugala+,van2018geometry}. GENERIC (General Equation for Non-Equilibrium Reversible–Irreversible Coupling)—that is, the unification of Hamiltonian and dissipative dynamics within a single geometric framework—provides a powerful toolset for analyzing nonequilibrium thermodynamics from both theoretical and practical perspectives. For foundational works on GENERIC, we refer to \cite{grmela1984bracket,grmela1984bracketb,grpd}, while several recent developments and extensions can be found in \cite{grmela2018generic,go,hco,og,pkg}.

Our overarching aim is to deepen the geometric understanding of GENERIC within the framework of contact geometry (see \cite{grmela-contact}). In pursuit of this goal, the present work revisits and extends two central problems previously studied in our earlier research—namely, the formulation of contact kinetic dynamics and the treatment of GENERIC within the geometric Hamilton–Jacobi framework. Let us comment on these issues one by one.

\textbf{Contact Kinetic Dynamics.}  
The Hamiltonian/geometric analysis of the kinetic formulation of a continuum starts from a group of diffeomorphisms that preserve particle motion and derives the dynamics in terms of momentum and/or density variables residing in the dual of the corresponding Lie algebra. 
Technically, this construction gives rise to Lie–Poisson dynamics \cite{MaRa99,holm2009geometric,liber87}. A geometric pathway for deriving such dynamics was introduced in the context of geometric kinetic theory in \cite{esgu11,esgu12}. In these works, the kinetic motion is obtained by starting from the particle-level dynamics and then applying lift operations together with jet (or vertical–holonomic) decompositions. This analysis was later extended to the study of the moments of the Poisson–Vlasov equations \cite{esgrmigu19,essu21}, and further generalized to locally conformally symplectic dynamics \cite{EsGeGu24} and time-dependent kinetic dynamics \cite{AtEsLeSa}. In that project, presented in \cite{EsGeGrGuPaSu,esgu11,esgu12}, an analysis was carried out to obtain a kinetic formulation for ensembles of particles evolving according to contact Hamiltonian vector fields. These studies thus provided the first versions of contact kinetic dynamics.

The first goal of the present paper is to develop a more comprehensive formulation of contact kinetic dynamics by making explicit certain structural and algebraic features—in particular, the gauge invariance associated with conformal factors—that were not addressed in earlier treatments. The novel formulation we propose couples the evolution of the density (distribution) function with that of an additional variable governing the evolution of the continuum volume. As we demonstrate in the main body of the paper, this coupling provides a new perspective on entropy, which can now be interpreted as an external variable within contact kinetic theory.

\textbf{Geometric Hamilton–Jacobi Realization of GENERIC.} 
The geometric Hamilton–Jacobi theory \cite{carihj} is a powerful framework that geometrically characterizes the classical Hamilton–Jacobi theory, offering a deeper geometric understanding of physical systems that admit Hamiltonian or Lagrangian representations. In recent years, numerous studies have explored various aspects and applications of the geometric Hamilton–Jacobi equation across a wide range of physical and mathematical contexts. For recent comprehensive reviews addressing this diversity, we refer to \cite{esen2022reviewing,roman2021overview}.

Our specific interest in the present work concerns the geometric Hamilton–Jacobi theory in the setting of contact geometry, which has been studied, for instance, in \cite{GraCont,de2021hamilton,leonsar2,cannarsa2019herglotz,WaYaZh}.
On a contact manifold, two types of dynamical vector fields naturally arise. The first is the evolution (contact) Hamiltonian vector field, defined with reference to the almost–Poisson structure on the contact manifold (see \cite{simoes-thermo}). The second is the conventional (or classical) contact Hamiltonian vector field, which is defined with respect to the Jacobi structure on the contact manifold.

In our earlier works \cite{esgrpa22a,esgrpa22b}, we developed a geometric realization of the GENERIC framework within contact Hamilton–Jacobi theory. In that formulation, GENERIC was obtained as the projection of an evolution contact Hamiltonian dynamics, and it was shown that GENERIC can be lifted to an evolution contact Hamiltonian dynamics provided that the entropy satisfies an associated Hamilton–Jacobi equation. However, evolution contact dynamics is not as natural as pure contact Hamiltonian dynamics, as it lacks the full geometric consistency of the Jacobi structure (see \cite{GraCont}). For this reason, in the present work we revisit the GENERIC framework within contact Hamilton–Jacobi theory once more. Here, GENERIC is derived as the projection of pure contact Hamiltonian dynamics, thereby providing a more geometrically natural formulation. Furthermore, unlike in our previous works \cite{esgrpa22a,esgrpa22b}, the present construction also incorporates the dynamics of the dissipation potential, which was previously absent.

To achieve this, we introduce a novel intermediate level—a specifically constructed graph space—within the geometric Hamilton–Jacobi framework for contact dynamics. We then reformulate the GENERIC system on this intermediate level, obtaining a fully contact-geometric representation of GENERIC, where the contact Hamiltonian system is coupled to the evolution of the dissipation potential itself. A key advantage of employing contact vector fields rather than evolution vector fields is that thermodynamics can now be recast as dynamics on manifolds of submanifolds, providing a unified geometric perspective on both reversible and irreversible processes.

One of the strengths of the contact geometric Hamilton–Jacobi framework of GENERIC is its ability to accommodate additional dynamical factors that naturally arise from the physical structure of the system under consideration. In this paper, we also demonstrate how microturbulence can be incorporated into the GENERIC framework, and how this extension can be geometrically represented within the contact geometric Hamilton–Jacobi theory.

\noindent \textbf{The content.} The present paper is organized into three main sections. The following section reviews symplectic, Poisson, and contact manifolds together with their corresponding Hamiltonian dynamics, in order to fix notation and complete the geometric preliminaries of the study. The section also presents cotangent and contact lifts, both for future reference and as illustrative applications.

In Section~\ref{sec.irr}, we begin with a review of geometric kinetic dynamics, providing a Hamiltonian (more precisely, Lie–Poisson) formulation of the Liouville (Vlasov) equations. Subsection~\ref{sec:gr} presents an algebraic analysis of contact diffeomorphisms, showing that the group of contact diffeomorphisms admits a semi-direct product structure incorporating conformal factors intrinsic to contact geometry. The corresponding Lie algebra, its dual, and the resulting Lie–Poisson dynamics are derived in detail.
In Subsection~\ref{sec:Ham-mom}, the space of contact Hamiltonian vector fields is identified with the space of infinitesimal contact transformations and analyzed as a subalgebra of the semi-direct product Lie algebra obtained in the previous subsection. Its dual is characterized, leading to a new formulation of contact kinetic dynamics expressed in terms of momenta. This novel model extends earlier formulations by incorporating the dynamics of volume change into the system.
Finally, Subsection~\ref{sec:Ham-den} reformulates the resulting contact kinetic dynamics in terms of density functions rather than momenta.

Section~\ref{sec.app} is devoted to the study of GENERIC within the contact Hamilton–Jacobi framework. Subsection~\ref{sec.HJ} provides a concise introduction to the contact Hamilton–Jacobi theory and introduces a new intermediate level—referred to as the graph space—within this setting. Subsection~\ref{subsecBW} develops a reformulation of GENERIC on this intermediate level, clarifying its geometric relation with pure contact Hamiltonian dynamics. As an application of these constructions, Subsection~\ref{sec:micc} demonstrates how microturbulence can be incorporated into GENERIC within this geometric framework.

The paper concludes with a discussion outlining several possible directions for future research.

\noindent \textbf{Notation.} In this work, we employ both index notation (assuming the Einstein summation convention) and vector--matrix notation using matrix and dot products. 
For a manifold with local coordinates $\mathbf{x} = (x^a)$, a vector field $X$ can be written as 
\begin{equation}
    X = X^a \, \partial_{x^a},
    \qquad \text{or equivalently} \qquad
    X = \mathbf{X} \cdot \nabla_{\mathbf{x}},
\end{equation}
where $\mathbf{X}$ denotes the column vector of components $(X^a)$, and $\nabla_{\mathbf{x}}$ represents the column vector of differential operators $(\partial_{x^a})$.

Similarly, a differential one-form $\alpha$ can be expressed as 
\begin{equation}
    \alpha = \alpha_a \, d x^a,
    \qquad \text{or equivalently} \qquad
    \alpha = \boldsymbol{\alpha} \cdot d\mathbf{x},
\end{equation}
where $\boldsymbol{\alpha}$ denotes the row vector of covariant components $(\alpha_a)$, and $d\mathbf{x}$ stands for the column vector of differentials $(d x^a)$.

Thus, $\mathbf{X}$ and $\boldsymbol{\alpha}$ are understood as contravariant and covariant representations, respectively, and the pairing $\langle \alpha, X \rangle = \boldsymbol{\alpha} \cdot \mathbf{X} = \alpha_a X^a$ holds under the Einstein summation convention.

\section{Particle  Dynamics}
\subsection{Dynamics on Poisson Manifolds} \label{sec.rev}

Macroscopic systems are composed of $\sim 10^{23}$ microscopic particles. In classical mechanics, the time evolution of the particles is governed by Hamilton's equations in the Poisson framework. Let us recall this geometrically, following \cite{abrahammarsden,holm2009geometric,liber87, MaRa99}. 

\noindent \textbf{(Almost) Poisson Manifold.}
A skew-symmetric $\mathbb{R}$-bilinear bracket, defined on the space $\mathcal{C}^{\infty}(P)$ of a manifold $P$,
\begin{equation}\label{PoissonBracket}
  \{\bullet,\bullet\}:\mathcal{C}^{\infty}(P) \times \mathcal{C}^{\infty}(P)\longrightarrow \mathcal{C}^{\infty}(P).
\end{equation}
is an almost Poisson bracket if it satisfies the Leibniz identity:  
\begin{equation}\label{LeibId}
 \{A,B\cdot C\}= \{A,B\}\cdot C+ B \cdot \{A,C\},
\end{equation}
for all $A$, $B$ and $C$ in $\mathcal{C}^{\infty}(P)$.  We identify an almost Poisson bracket $\{\bullet,\bullet\}$ with a bivector field $\mathbb{L}$ on $P$ according to the following definition 
\begin{equation} \label{bivec-PoissonBra}
\mathbb{L}(dA,dB)=\{A,B\}.
\end{equation}    
A manifold equipped with an almost Poisson bracket is called an almost Poisson manifold. 

An almost Poisson bracket turns out to be a Poisson bracket if it additionally satisfies the Jacobi identity:
\begin{equation}\label{Jac-ident-Poiss}
\circlearrowright \{A,\{B,C\}\}=\{A,\{B,C\}\}+\{B,\{C,A\}\}+\{C,\{A,B\}\}=0,
\end{equation}
for all $A$, $B$ and $C$ in $\mathcal{C}^{\infty}(P)$.   
A bracket defined through a bivector field $\mathbb{L}$, as in \eqref{bivec-PoissonBra}, satisfies the Jacobi identity if and only if $\mathbb{L}$ commutes with itself under the Schouten–Nijenhuis bracket:  
\begin{equation}\label{Poisson-cond}
[\mathbb{L},\mathbb{L}]_{SN} = 0 .
\end{equation}

\noindent \textbf{Hamiltonian Dynamics on (Almost) Poisson Framework.} Consider an almost Poisson manifold $(P,\{\bullet,\bullet\})$. Given a Hamiltonian (energy) function $E$, the Hamiltonian vector field $X_E$ is defined in terms of the almost Poisson bracket as
\begin{equation}\label{Hamvf-}
X_E(\mathbf{x})=\{\mathbf{x}, E\},
\end{equation}
for $\mathbf{x}$ (which turns out to be the state variable) in $P$. Accordingly, the Hamiltonian dynamics is defined as
\begin{equation} \label{HamEq}
\dot{\mathbf{x}}=\{\mathbf{x},E\}, \qquad \text{or equivalently}\quad \dot{x}^a=\{x^a,E\},
\end{equation}  
where we have assumed local coordinates $\mathbf{x}=(x^a)$. The skew-symmetry of the bracket manifests that the Hamiltonian function $E$ is conserved all along the motion that is $\dot{E}=0$. This physically corresponds to the conservation of energy. 

Referring to \eqref{bivec-PoissonBra}, in local coordinates $\mathbf{u}=(u^i)$, we have the Poisson bivector field $\mathbb{L}=\mathbb{L}^{ij} \partial_{u^i}\wedge \partial_{u^j}$, hence the Hamilton's equations \eqref{HamEq} can also be written as 
\begin{equation}\label{equ1}
\dot{\mathbf{u}}=\mathbb{L} E_\mathbf{u}, \qquad \text{or equivalently}\quad
\dot{u}^i=\mathbb{L}^{ij}\frac{\partial E} {\partial x^j}. 
\end{equation} 

\noindent\textbf{Lie–Poisson Dynamics.} 
In this paper, we are interested in a particular class of Poisson manifolds called Lie–Poisson manifolds, \cite{MaRa99,holm2009geometric}.  
We begin with a Lie group $G$. In the present work, the Lie group of interest will be a diffeomorphism group.  
The tangent space at the identity element of $G$ forms a Lie algebra, denoted by $\mathfrak{g}$.  
The Lie bracket, also referred to as the adjoint representation, is given by
\begin{equation}
    [\bullet,\bullet]: \mathfrak{g} \times \mathfrak{g}\longrightarrow \mathfrak{g},\qquad (\xi,\eta)\mapsto ad_\xi(\eta)=[\xi,\eta]. 
\end{equation}
We denote the dual space of the Lie algebra $\mathfrak{g}$ by $\mathfrak{g}^*$.  
Dualizing the adjoint action yields the coadjoint action of $\mathfrak{g}$ on $\mathfrak{g}^*$:
\begin{equation}
    ad^*: \mathfrak{g} \times \mathfrak{g}^*\longrightarrow \mathfrak{g}^*,\qquad (\xi,\nu)\mapsto ad^*_\xi \nu,
\end{equation}
defined, for all $\xi$ and $\eta$ in $\mathfrak{g}$ and $\nu$ in $\mathfrak{g}^*$, by
\begin{equation}\label{coad}
    \langle ad^*_\xi \nu, \eta \rangle = \langle \nu, ad_\xi \eta \rangle, \qquad \forall \, \eta \in \mathfrak{g}.
\end{equation}

For two real-valued functionals $A$ and $B$ defined on $\mathfrak{g}^*$,  
the (plus) Lie–Poisson bracket is given as 
\begin{equation}\label{LP-bra}
    \{A,B\}^{\mathrm{LP}}(\nu)=\Big \langle \nu, \Big[\frac{\delta A}{\delta \nu},\frac{\delta B}{\delta \nu}\Big] \Big \rangle,
\end{equation}
where the terms $\delta A/\delta \nu$ and $\delta B/\delta \nu$ denote the variational derivatives  
(Fréchet derivatives in the infinite-dimensional case, partial derivatives in finite dimensions).  
The pairing $\langle \bullet, \bullet \rangle$ is that between $\mathfrak{g}^*$ and $\mathfrak{g}$. The Lie-Poisson space is the Poisson manifold determined by the two-tuple $(\mathfrak{g}^*,\{\bullet,\bullet\}^{\mathrm{LP}})$. 

To derive the equation of motion, let $H$ be defined on  $\mathfrak{g}^*$ be a Hamiltonian functional.  
Then the dynamics of an observable $A$ is governed by Hamilton’s equation

To derive the equation of motion, let $\mathcal{H}$ be a Hamiltonian function(al) defined on the dual space $mathfrak{g}^* \to \mathbb{R}$.   
Then the dynamics of an observable $A$ is given by the Hamilton's equation
\begin{equation}
   \frac{dA}{dt} = \{A,\mathcal{H}\}^{\mathrm{LP}}(\nu).
\end{equation}
The left-hand side evaluates to
\begin{equation}\label{fat1}
  \frac{dA}{dt} =  \Big\langle \frac{\delta A}{\delta \nu}, \frac{d\nu}{dt} \Big\rangle,
\end{equation}
while the right-hand side reads
\begin{equation}\label{fat2}
\begin{split}
\{A,\mathcal{H}\}^{\mathrm{LP}}(\nu) 
&=\Big \langle \nu, \Big[\frac{\delta A}{\delta \nu},\frac{\delta \mathcal{H}}{\delta \nu}\Big] \Big \rangle \\ 
&= - \Big \langle \nu, \Big[\frac{\delta \mathcal{H}}{\delta \nu},\frac{\delta A}{\delta \nu}\Big] \Big \rangle \\
&=   - \Big \langle \nu, ad_{\delta \mathcal{H}/\delta \nu}  \frac{\delta A}{\delta \nu}  \Big \rangle \\
&=  -  \Big \langle ad_{\delta \mathcal{H}/\delta \nu}^*\nu, \frac{\delta A}{\delta \nu}  \Big \rangle,
\end{split}
\end{equation}
where in the second equality we used the skew-symmetry of the Lie algebra bracket,  
in the third we identified the adjoint action with the Lie bracket,  
and in the last line, using \eqref{coad}, we obtain the coadjoint action.  
Comparing \eqref{fat1} and \eqref{fat2}, we obtain the Lie–Poisson equations, also known as the coadjoint flow,
\begin{equation}\label{LP-dyn}
    \frac{d\nu}{dt} = - ad_{\delta \mathcal{H}/\delta \nu}^*\nu.
\end{equation}
This itinerary—from a Lie group to its Lie algebra, then to its dual, and finally to the coadjoint flow on $\mathfrak{g}^*$—is fundamental for the dynamics considered in the present work.

\noindent \textbf{Casimirs and Degeneracy.} A Casimir function $C$ on an almost Poisson manifold is a function that commutes with all smooth functions under the almost Poisson bracket, that is, $\{C,A\} = 0$ for all functions $A$. As a result, the Hamiltonian vector field associated with $C$ vanishes, $X_C = 0$. An almost Poisson manifold that admits no non-trivial (i.e., non-constant) Casimir functions is called an almost symplectic manifold. For Poisson manifolds, the notion of a Casimir function is defined in the same way, and similarly, a Poisson manifold with no non-trivial Casimir functions is called a symplectic manifold. 

Having no non-trivial Casimir functions is termed non-degeneracy, which implies that a Poisson manifold with no non-trivial Casimir functions—that is, a symplectic manifold—must be even-dimensional. In what follows, we restrict our attention to symplectic manifolds.

\noindent \textbf{Symplectic Manifolds and Dynamics.} 
A $2n$-dimensional symplectic manifold $P$  admits a closed (that is $d\Omega=0$) non-degenerate two-form $\Omega$. 
We denote a symplectic manifold by a pair $(P,\Omega)$. The non-degeneracy can be rephrased by arguing that  $\Omega^n\neq 0$. This leads to the definition of symplectic volume 
\begin{equation}\label{vol-symp}
    \mathrm{Vol}_{\Omega}=\Omega^n. 
\end{equation}
For a given Hamiltonian function $E$ on $(P,\Omega)$, the Hamiltonian vector field $X_E$ determines the Hamiltonian dynamics:
\begin{equation} \label{HamEq.symp}
\iota_{X_E}\Omega = dE. 
\end{equation}
Referring to the symplectic two-form $\Omega$, one can define a non-degenerate Poisson bracket as
\begin{equation}\label{poisson-on-ham}
\{A,B\} = \Omega(X_A, X_B).
\end{equation}
Here, $X_A$ and $X_B$ are the Hamiltonian vector fields determined by Hamilton's equation \eqref{HamEq.symp}. Thus, as claimed earlier, symplectic manifolds carry a natural Poisson structure.

Cotangent bundles are fundamental examples of symplectic manifolds by carrying the canonical symplectic two-form. 
Accordingly, we define the canonical (Liouville) one-form $\Theta_{M}$ \index{canonical (Liouville) one-form} on $T^{\ast }M$ as 
\begin{equation}
\Theta_{M}( X) =\left\langle \tau
_{T^{\ast }M}\left( X \right) ,T\pi _{M}\left( X
\right) \right\rangle ,  \label{canonicaloneform}
\end{equation}
for a vector field $X$ on $T^{\ast }M$. Here, we used the tangent bundle projection $\tau
_{T^{\ast }M}: TT^*M\mapsto T^*M$ and the tangent mapping $T\pi _{M}:  TT^*M\mapsto T^*M$ of the cotangent bundle projection $\pi_M$. We present the following diagram to show these mappings
\begin{equation}
\xymatrix{
&TT^{\ast }M\ar[dl]^{T\pi _{M}}\ar[dr]_{\tau _{T^{\ast }M}}
\\TM \ar[dr]_{\tau_{M}}&&T^*M\ar[dl]^{\pi_{M}}\ar@/_1pc/[ul]_{X}
\\&M } \label{T}
\end{equation} 
as well as the vector field $X$.  Minus of the exterior derivative of the canonical one-form $\Theta_{M}$ is the canonical symplectic two-form
\begin{equation}\label{omega-can}
\Omega_{M}=-d\Theta_{M}. 
\end{equation}
We denote this canonical symplectic structure by $(T^*M,\Omega_M)$. 


In the symplectic picture of $T^*M$ with Darboux coordinates $(\mathbf{x},\mathbf{x}^*) = (x^a, x_a^*)$, the canonical one-form is
\begin{equation}
    \Theta_M = \mathbf{x}^* \cdot d\mathbf{x}, \qquad \text{or equivalently} \qquad \Theta_M = x_a^*\, dx^a,
\end{equation}
while the symplectic two-form is the constant two-form
\begin{equation}
    \Omega_M = d\mathbf{x} \wedge d\mathbf{x}^*, \qquad \text{or equivalently} \qquad \Omega_M = dx^a \wedge dx_a^*.
\end{equation}

In terms of the matrix notation, we have the canonical symplectic two-form and canonical Poisson bivector field as
\begin{equation}
 \Omega_M=\mathbb{L}^{-1}=\left(\begin{array}{cc}0&-\mathbb{I}\\\mathbb{I}&0\end{array}\right),
\end{equation}
where $\mathbb{I}$ is the identity matrix. This reads the  bivector $\mathbb{L}$ as
\begin{equation}\label{AP-Bra}
\mathbb{L}= \nabla_\mathbf{x}\wedge \nabla_{\mathbf{x}^*}, \qquad \text{or equivalently} \quad  \mathbb{L}=
\frac{\partial  }{\partial x^{a}}\wedge \frac{\partial  }{\partial x^*_{a}}. 
\end{equation} 
Hence, we write the canonical Poisson bracket \eqref{poisson-on-ham} as
\begin{equation}\label{canPoibra}
\begin{split}
\{A,B\} &=(\nabla_\mathbf{x} A,\nabla_{\mathbf{x}^*} A)\left(\begin{array}{cc}0&\mathbb{I}\\-\mathbb{I}&0\end{array}\right)\left(\begin{array}{cc} \nabla_\mathbf{x} B \\\nabla_{\mathbf{x}^*} B \end{array}\right) \\ 
    &=  \nabla_\mathbf{x} A \cdot \nabla_{\mathbf{x}^*}B- \nabla_{\mathbf{x}^*}A  \cdot \nabla_{\mathbf{x}}B. 
\end{split}
\end{equation}
Hence, in Darboux coordinates $(\mathbf{x},\mathbf{x}^*)=(x^a,x^*_a)$, for a Hamiltonian function $E$, the Hamilton's equations are computed to be
\begin{equation}\label{HamEq-sym}
  \dot{\mathbf{x}}
  = \nabla_{\mathbf{x}^*} E, \qquad  \dot{\mathbf{x}}^*= - \nabla_\mathbf{x} E. 
\end{equation}

\subsection{Dynamics on Contact Manifolds} \label{sec:contact-dyn}

In odd dimensions, a symplectic structure cannot exist since a symplectic two-form is necessarily non-degenerate and hence requires the manifold to be even-dimensional. Instead, one encounters contact structures, which may be thought of as the odd-dimensional counterpart of symplectic geometry. However, contact structures do not define a Poisson structure, as the corresponding bracket operation fails to satisfy the Leibniz identity and, in general, the Hamiltonian function is not conserved. These structural differences highlight fundamental distinctions between symplectic and contact geometries. Let us now turn to examination of contact structures.

\noindent \textbf{Contact Manifolds.} 
Let $\mathbb{M}$ be an odd-dimensional manifold, say of dimension $2n + 1$. A contact structure on $\mathbb{M}$ is a maximally non-integrable smooth distribution of codimension one \cite{Arn,liber87}. In this paper, to define a contact structure, we assume the existence of a (global) contact one-form $\eta$ satisfying the non-integrability (or non-degeneracy) condition
\begin{equation}\label{non-int}
\eta \wedge (d\eta)^n \neq 0.
\end{equation}
This condition corresponds to requiring that $\mathbb{M}$ is coorientable by determining the contact volume form 
\begin{equation}\label{vol-cont}
\mathrm{Vol}_\eta= \eta \wedge (d\eta)^n.
\end{equation}
As a consequence, there exists a distinguished vector field, called the Reeb vector field and denoted by $\mathcal{R}$, which is uniquely defined through
\begin{equation}\label{Reeb}
\iota_{\mathcal{R}} \eta = 1, \quad \iota_{\mathcal{R}} d\eta = 0.
\end{equation}
The Reeb vector field generates the flow along the direction transverse to the contact distribution and plays a role analogous to, but distinct from, Hamiltonian vector fields in symplectic geometry, as it is not generally associated with the conservation of a Hamiltonian function.

The non-degeneracy condition \eqref{non-int} leads us to define the musical flat isomorphism from the space $\mathfrak{X}(\mathbb{M})$ of vector fields on $\mathbb{M}$ to the space $\Lambda^1(\mathbb{M})$ of one-forms on $\mathbb{M}$: 
\begin{equation}\label{flat-map}
\flat:\mathfrak{X}(\mathbb{M})\longrightarrow \Lambda^1(\mathbb{M}),\qquad X\mapsto \iota_Xd\eta+\eta(X)\eta.
\end{equation} 
Being an isomorphism, the musical flat mapping admits an inverse operation from $\Lambda^1(\mathbb{M})$ to $\mathfrak{X}(\mathbb{M})$. 
We denote this inverse by the sharp mapping $\sharp$:
\begin{equation}\label{sharp}
\sharp:\Lambda^1(\mathbb{M}) \longrightarrow \mathfrak{X}(\mathbb{M}), \qquad \sharp = \flat^{-1}.
\end{equation}
Notice that the image of the Reeb vector field $\mathcal{R}$ under the flat mapping $\flat$ is precisely the contact one-form $\eta$, whereas the image of $\eta$ under the sharp mapping $\sharp$ is $\mathcal{R}$; that is,
\begin{equation}
\flat(\mathcal{R}) = \eta, \qquad \sharp(\eta) = \mathcal{R}.
\end{equation}
This observation highlights the ubiquitous role of the Reeb vector field in contact geometry.

Using the sharp mapping $\sharp$, we define the contact bivector field $\Lambda$ as
\begin{equation}\label{Lambda}
\Lambda(\alpha,\beta) = d\eta(\sharp\alpha, \sharp\beta).
\end{equation}
By contracting the contact bivector field $\Lambda$ from the right, we obtain another sharp mapping
\begin{equation}\label{Sharp-Delta}
\sharp_\Lambda: \Lambda^1(\mathbb{M}) \longrightarrow \mathfrak{X}(\mathbb{M}), \qquad 
\alpha \mapsto \Lambda(\bullet,\alpha) = \sharp \alpha - \alpha(\mathcal{R})\,\mathcal{R}.
\end{equation}

In terms of the contact bivector field $\Lambda$ and the Reeb vector field $\mathcal{R}$, the contact bracket of smooth functions on $\mathbb{M}$ is defined as
\begin{equation}\label{cont-bracket-L} 
\{A,B\}^{(C)} = \Lambda(dA,dB) + A\,\mathcal{R}(B) - B\,\mathcal{R}(A).
\end{equation}
The contact bracket satisfies the Jacobi identity but generally fails to satisfy the Leibniz rule. This property reflects the Jacobi structure of contact manifolds. More precisely, a contact manifold $(\mathbb{M},\eta)$ is a Jacobi manifold with the pair $(\Lambda,\mathcal{R})$, consisting of the contact bivector field and the Reeb vector field \cite{Lichnerowicz-Jacobi,Marle-Jacobi}. 
The extra terms $A\,\mathcal{R}(B) - B\,\mathcal{R}(A)$ in the contact bracket account for the flow along the Reeb vector field and distinguish the contact bracket from the usual Poisson bracket. They reflect the fact that, in contact geometry, the Hamiltonian function is generally not conserved and the dynamics includes a contribution transverse to the contact distribution.

\noindent
\textbf{Extended Cotangent Bundle.} 
An example of a coorientable contact manifold can be constructed as follows. Consider a trivial line bundle over a manifold $M$. The first jet of this bundle is diffeomorphic to the product space $T^*M \times \mathbb{R}$, which we call the extended cotangent bundle. This space naturally admits a (global) contact one-form.  
If, instead, one chooses a line bundle that is not necessarily trivial, then the resulting contact structure is defined in terms of a distribution, but it does not necessarily admit a global contact one-form. 

On the extended cotangent bundle, by pulling back the canonical one-form $\Theta_{M}$ on $T^*M$ and the exact one-form $dz$ on $\mathbb{R}$, we define a contact one-form
\begin{equation}\label{eta-Q}
\eta_M = dz - \Theta_{M}.
\end{equation}
Indeed, we have that the one-form $\eta$ in \eqref{eta-Q} satisfies the non-integrability  condition
\eqref{non-int}. We denote this contact manifold by $(T^*M \times \mathbb{R},\eta_M)$. In this realization, $T^*M \times \mathbb{R}$ is called contactification of the symplectic manifold $T^*M$. 

The existence of Darboux coordinates $( \mathbf{x}, \mathbf{x}^*)=(x^a, x_a^*)$ on the cotangent bundle $T^*M$ allows us to introduce Darboux coordinates $( \mathbf{x}, \mathbf{x}^*,z)=(x^a, x_a^*, z)$ on the contact manifold $T^*M \times \mathbb{R}$. In these coordinates, the contact one-form reads
\begin{equation}\label{can-contat-form}
\eta_M = dz - \mathbf{x}^* \cdot d\mathbf{x}, \qquad \text{or equivalently} \qquad \eta_M = dz - x_a^*\, dx^a.
\end{equation}
The corresponding Reeb vector field is simply
\begin{equation} 
\mathcal{R} = \nabla_z.
\end{equation}
The musical isomorphism $\flat$ in \eqref{flat-map} and $\sharp$ in \eqref{sharp} are computed to be
\begin{equation}\label{loc-flat-sharp}
    \begin{split}
      \flat\big(\mathbf{X}_{\mathbf{x}}\cdot \nabla_\mathbf{x} + \mathbf{X}_{\mathbf{x}^*}\cdot \nabla_{\mathbf{x}^*} + X_{z} \nabla_z \big) = &  - \big(\mathbf{X}_{\mathbf{x}^*}+ \mathbf{x}^*X_{z}-(\mathbf{X}_{\mathbf{x}}\cdot \mathbf{x}^*) \mathbf{x}^*\big) \cdot d \mathbf{x} \\ 
      & \qquad +\mathbf{X}_{\mathbf{x}}\cdot d \mathbf{x}^* + \big(X_{z}-  \mathbf{x}^*\cdot \mathbf{X}_{\mathbf{x}}\big) dz,
      \\\sharp(\Pi_{\mathbf{x}} \cdot  d\mathbf{x} + \Pi_{\mathbf{x}^*}\cdot d\mathbf{x}^* + \Pi_z dz) = ~& \Pi_{\mathbf{x}^*}\cdot \nabla_\mathbf{x} - \big(\Pi_{\mathbf{x}}+ \Pi_z \mathbf{x}^* \big)\cdot \nabla_{\mathbf{x}^*} \\& \qquad + \big( \Pi_z+ \Pi_{\mathbf{x}^*}\cdot \mathbf{x}^*\big)\nabla_z,
    \end{split}
\end{equation}
respectively. In this notation, the contact bivector field $\Lambda$ defined in \eqref{Lambda} turns out to be 
\begin{equation}\label{AP-Bra+}
\Lambda 
= \nabla_\mathbf{x} \wedge \nabla_{\mathbf{x}^*} - \mathbf{x}^* \cdot \nabla_{\mathbf{x}^*} \wedge \nabla_z
\end{equation}
while the contraction of this bivector field by a one-form $\Pi$ that is the musical mapping $\sharp_\Lambda
$ in  \eqref{Sharp-Delta} is 
\begin{equation}\label{zap}
\sharp_{\Lambda}(\Pi_{\mathbf{x}} \cdot  d\mathbf{x} + \Pi_{\mathbf{x}^*}\cdot d\mathbf{x}^* + \Pi_z dz) = \Pi_{\mathbf{x}^*}\cdot \nabla_\mathbf{x} - \big(\Pi_{\mathbf{x}}+ \Pi_z \mathbf{x}^* \big)\cdot \nabla_{\mathbf{x}^*}  + \big(  \Pi_{\mathbf{x}^*}\cdot \mathbf{x}^*\big)\nabla_z.
\end{equation}
Hence, in this local picture, the contact bracket \eqref{cont-bracket-L} takes the form
 \begin{equation}\label{Lag-Bra-}
\begin{split}
\{A,B\}^{(C)} &= \nabla_\mathbf{x} A \cdot \nabla _{\mathbf{x}^*}B - \nabla_{\mathbf{x}^*} A \cdot \nabla_\mathbf{x} B 
+ \big(A - \mathbf{x}^* \cdot \nabla_{\mathbf{x}^*}A\big) \nabla_z B\\ & \quad - \big(B - \mathbf{x}^* \cdot \nabla_{\mathbf{x}^*} B\big) \nabla_z A.
\end{split}
\end{equation}

 \noindent \textbf{Legendrian Submanifolds.}
Let $(\mathbb{M},\eta)$ be a contact manifold. Recall the associated bivector field $\Lambda$ defined in \eqref{Lambda}. Consider a linear subbundle $\mathbb{S}$ of the tangent bundle $T\mathbb{M}$ (that is, a distribution on $\mathbb{M}$). We define the contact complement of $\mathbb{S}$ as
\begin{equation}
\mathbb{S}^\perp  = \sharp_ \Lambda(\mathbb{S}^o),
\end{equation}
where the sharp map on the right-hand side is the one in   \eqref{Sharp-Delta} and   $\mathbb{S}^o$ is the annihilator of  $\mathbb{S}$. 

A submanifold is said to be a Legendrian submanifold  if 
\begin{equation}
TS= {TS}^{\perp }.
\end{equation}
One can easily prove that a submanifold $S$ of $\mathbb{M}$ is Legendrian if and only if it is a maximal integral manifold of $\ker \eta$. In this case, the dimension of $S$ must be $n$ (see \cite{lela19,liber87}). So, we may claim that a Legendrian submanifold $S$ can be seen as a submanifold of $\mathbb{M}$ with dimension $n$ such that $\eta|_{S}=0$. 

Consider the contact manifold $(T^*M \times \mathbb{R},\eta_M)$. 
The first jet prolongation of a real-valued function $A$ on the base manifold $M$ is a section
\begin{equation}\label{j1F}
j^1 A: M \longrightarrow  T^* M \times \mathbb{R},\qquad \mathbf{x}\mapsto (\mathbf{x},\nabla_\mathbf{x} A,A(\mathbf{x})).
\end{equation}
The image space is a Legendrian submanifold of $T^* M \times \mathbb{R}$ defined as 
\begin{equation}
S=\mathrm{im}(j^1 A)=\big\{\big(\mathbf{x},\nabla_\mathbf{x} A(\mathbf{x}),A(\mathbf{x})\big) \in  T^* M \times \mathbb{R} : A \in C^\infty (M) \big\}.
\end{equation}
The converse of this assertion is also true, that is, if the image space of a section of $T^*M \times \mathbb{R} \mapsto M$ is a Legendrian submanifold, then it is the first prolongation of a function $F$.  

This can be seen in equilibrium thermodynamics through the fundamental thermodynamic relation \cite{callen,Gibbs,hermann}.  
Consider the base coordinates $(T,N,V)$, where $T$ denotes the temperature, $N$ the number of moles, and $V$ the volume.  
The corresponding conjugate variables are $(S,\mu,P)$, where $S$ is the entropy, $\mu$ the chemical potential, and $P$ the pressure.  
With the one-dimensional extension given by $U$, the internal energy, the associated contact one-form is
\begin{equation}
  \eta = dU - T\, dS + \mu\, dN + P\, dV, 
  \qquad \eta|_{\mathcal{N}} = 0 .
\end{equation}
Here, the Legendrian submanifold $\mathcal{N}$ is determined by the fundamental thermodynamic relation
\[
U = U(S,V,N).
\]

There is a more general way to define Legendrian submanifolds of $T^*M \times \mathbb{R}$ by allowing the generating function $A$ to depend on auxiliary variables as well.  
Geometrically, this corresponds to defining $A$ not only on the base manifold $M$, but on the total space of a fiber bundle structure over $M$.  

Let $(\mathcal{B}, \tau, M)$ be a fiber bundle with local coordinates $\mathbf{x}$ on the base and $(\mathbf{x}, \mathbf{c})$ on the total space.  
Assume that $A = A(\mathbf{x}, \mathbf{c})$. If the rank of the Hessian-type matrix composed of the second derivatives
\begin{equation}\label{MorseCon}
\left( A_{\mathbf{c}\mathbf{c}} \quad A_{\mathbf{x}\mathbf{c}} \right)
\end{equation}
is maximal, then the submanifold
\begin{equation}\label{MFGen-C}
S = \left\{ (\mathbf{x}, \nabla_\mathbf{x} A(\mathbf{x},\mathbf{c}), A(\mathbf{x},\mathbf{c})) \in T^*M \times \mathbb{R} 
:\; \nabla_\mathbf{c} A(\mathbf{x},\mathbf{c}) = 0 \right\}
\end{equation}
is a Legendrian submanifold, see \cite{esen2021contact,IbaLeonMarrDiego}.

 \noindent
\textbf{Contact Hamiltonian Vector Fields and Dynamics.}
Consider a contact manifold $(\mathbb{M},\eta)$. For a Hamiltonian function $H$ on $\mathbb{M}$, there is a corresponding contact Hamiltonian vector field 
$X_{H}$ given by
\begin{equation}\label{con-Ham-v-f}
\iota_{X_{H}}\eta =-H,\qquad \iota_{X_{H}}d\eta =dH-\mathcal{R}(H) \eta
\end{equation}
where $\mathcal{R}$ is the Reeb vector field, and $H$ is called a contact Hamiltonian function. We cite \cite{Br17,BrCrTa17,lela19,lela19,de2020review,de2021hamilton,esen2021contact} for some recent works on Hamiltonian dynamics on contact framework. In terms of the musical isomorphisms \eqref{flat-map} and \eqref{sharp}, we have that 
\begin{equation}\label{3-def2}
\begin{split}
    \flat(X_{H}) &=dH-(\mathcal{R}(H)+H)\eta,\\
    \sharp(dH) &= X_{H}+ (\mathcal{R}(H)+H) \mathcal{R},
\end{split} 
\end{equation}
respectively. We shall need these realizations in the sequel. 

A direct computation proves that the contact bracket and the contact vector field are related as
\begin{equation} \label{pelin}
\begin{split}
\{F,H\}^{(C)} &=- \iota_{[X_{H},X_{F}]}\eta =-\mathcal{L}_{X_{H}}\iota_{X_{F}} \eta+ 
\iota_{X_{F}}\mathcal{L}_{X_{H}} \eta
\\&=-\mathcal{L}_{X_{H}}(-F) + \iota_{X_{F}}(-\mathcal{R}(H)\eta)
\\&= X_{H}(F)+F\mathcal{R}(H).
\end{split} 
\end{equation}
The identity 
\begin{equation}\label{pelin2}
    [X_A,X_H]=-X_{\{A,H\}^{(C)} }
\end{equation}
reads the Jacobi identity for contact Hamiltonian vector fields. 

Referring to the Darboux coordinates $( \mathbf{x}, \mathbf{x}^*,z)$, for a Hamiltonian function $H=H( \mathbf{x}, \mathbf{x}^*,z)$, the contact Hamiltonian vector field is computed to be 
\begin{equation}
    X_H=\nabla_{\mathbf{x}^*} H\cdot \nabla_{\mathbf{x}} - (\nabla_{\mathbf{x}} H + \nabla_z H\mathbf{x}^*)\cdot \nabla_{\mathbf{x}^*} + (\mathbf{x}^*\cdot \nabla_{\mathbf{x}^*} H- H)  \nabla_{z}. 
\end{equation}
Therefore, the contact Hamiltonian
dynamics is the set of equations of motion:  
\begin{equation}\label{conham}
\begin{split}
\dot{\mathbf{x}}&=\nabla_{\mathbf{x}^*} H, \\ \dot{\mathbf{x}}^*&= -\nabla_{\mathbf{x}} H - \nabla_z H\mathbf{x}^*, \\   \dot{z} &= \mathbf{x}^*\cdot \nabla_{\mathbf{x}^*} H- H . 
\end{split}
\end{equation}

The contact Hamiltonian dynamics does preserves neither Hamiltonian function nor contact volume  $\mathrm{Vol}^{c}$ in \eqref{vol-cont}. For the former, we get 
\begin{equation}
{\mathcal{L}}_{X_{H}} \, H = - \mathcal{R}(H) H.
\end{equation}
Assuming the dimension of $\mathbb{M}$ to be $2n+1$, we compute  
\begin{equation}
{\mathcal{L}}_{X_{H}}  \, (\eta \wedge (d \eta)^n) = - (n+1) \, \mathcal{R} (H) \eta \wedge (d\eta)^n.
\end{equation}
Hence, a contact Hamiltonian vector field is not necessarily divergence-free, instead the divergence is obtained as
\begin{equation} \label{div-X-H}
\mathrm{div}(X_{H})= -  (n+1)  \mathcal{R} (H).
\end{equation}
 In Darboux coordinates, the divergence of a contact Hamiltonian vector field \eqref{div-X-H} is
\begin{equation} \label{div-X-H-}
\mathrm{div}(X _{H})=  - (n+1) \frac{\partial H}{\partial z}.
\end{equation}
However, for a nowhere vanishing Hamiltonian function $H$, the quantity 
\begin{equation}
{H}^{-(n+1)}  \eta \wedge (d\eta)^n 
\end{equation}
is preserved along the motion, see  \cite{BrLeMaPa20}.

\subsection{Symplectic and Contact Lifts of Particle Motion} \label{lpd}

 We present lifts of the particle motion on a manifold for both Poisson and contact frameworks. We begin with the complete cotangent lift and then extend it to the contact lift.

\noindent \textbf{Complete Cotangent Lift.} Consider a dynamics (given by a vector field) $X$ defined on a manifold $M$. There are numerous ways to lift $X$ to the cotangent bundle. The so-called complete cotangent lift of $X$ is a vector field $\hat{X}$ on the cotangent bundle $T^*M$ and given by
\begin{equation}\label{hatX}
T\pi _{M}\circ  \hat{X}=X\circ \pi _{M},
\end{equation}
where $T\pi _{M}$ is the tangent mapping of the cotangent bundle projection $\pi _{M}$ \cite{marsden-67, MaRa99,YaPa67}. Whatever the vector field $X$ is, the complete cotangent lift $\hat{X}$ is a Hamiltonian vector field on the symplectic manifold $T^*M$: 
\begin{equation}
    \iota_{\hat{X}}\Omega_M = d \hat{H}.
\end{equation}
Here, the Hamiltonian function is computed to be
\begin{equation}
    \hat{H}(\mathbf{x},\mathbf{x}^*)  = \mathbf{x}^* \cdot X(\mathbf{x}).
\end{equation}
Hence, the dynamics governed by the complete cotangent lift (that is the symplectic Hamiltonian motion) is  
\begin{equation}\label{ccl}
\dot{\mathbf{x}}= X(\mathbf{x}),\qquad  \dot{\mathbf{x}}^*= - \nabla_{\mathbf{x}}( \mathbf{x}^* \cdot X). 
\end{equation}
Notice that, this lift preserves the dynamics on the base variable $\mathbf{x}$.  

As a particular example for the complete cotangent lift, we assume now that the dynamics on the base manifold is Hamiltonian by itself. That is, $(M,\{\bullet,\bullet\})$ is a Poisson manifold with the Poisson bivector field $\mathbb{L}$, and $X=X_E=\mathbb{L}E_\mathbf{x}$ (see \eqref{equ1}) for a Hamiltonian (energy) function $E=E(\mathbf{x})$ defined on $M$. In this case, the complete cotangent lift $\hat{X}_E$ determines a Hamiltonian dynamics with respect to the canonical symplectic structure on the cotangent bundle $T^*M$, and the Hamiltonian function is 
\begin{equation}\label{E-T*M}
    \hat{E}(\mathbf{x},\mathbf{x}^*)=\mathbf{x}^* \cdot \mathbb{L}E_\mathbf{x}. 
\end{equation}
The lifted dynamics \eqref{ccl} takes the particular form:  
\begin{equation}\label{ccl-Ham}
\dot{\mathbf{x}}= \mathbb{L}E_\mathbf{x},\qquad  \dot{\mathbf{x}}^*= - \nabla_{\mathbf{x}} ( \mathbf{x}^* \cdot \mathbb{L}E_\mathbf{x}). 
\end{equation}
Apart from being Hamiltonian, the dynamics \eqref{ccl-Ham} has the following properties: (i) the Lagrangian submanifold obtained by the image of the exterior derivative of the Hamiltonian function $E$, that is 
\begin{equation}
\mathrm{im}(dE)=\{(\mathbf{x},\mathbf{x}^*)\in T^*M  : \mathbf{x}^*=E_\mathbf{x} \},
\end{equation}
is invariant, and (ii)  the dynamics \eqref{ccl-Ham} becomes  \eqref{equ1} on $\mathrm{im}(dE)$ .

\noindent \textbf{Complete Contact Lift.} Only a small step now brings mechanics to the framework of thermodynamics. We passed from \eqref{equ1} to (\ref{ccl-Ham}) by lifting the time evolution taking place in $M$ to the time evolution taking place in the cotangent bundle $T^*M$. We now make still another extension. We lift the time evolution from $T^*M$ to its contactification  $\mathbb{M}= T^*M\times \mathbb{R}$. 

Notice that there is a canonical inclusion of the cotangent bundle $T^*M$ into the product manifold $T^*M \times \mathbb{R}$. Referring to this inclusion and the Hamiltonian function $\hat{E}$ in \eqref{E-T*M-}, we define a Hamiltonian function on  $T^*M \times \mathbb{R}$ as
\begin{equation}\label{E-T*M-}
    \bar{E}(\mathbf{x},\mathbf{x}^*,z)=\mathbf{x}^* \cdot \mathbb{L}E_\mathbf{x}. 
\end{equation}
For this Hamiltonian function, referring to \eqref{con-Ham-v-f}, we determine the contact Hamiltonian vector field $X_{\bar{E}}$: 
\begin{equation}
\iota_{X_{\bar{E}}}\eta =-\bar{E},\qquad \iota_{X_{\bar{E}}}d\eta =d\bar{E}-\mathcal{R}(\bar{E}) \eta,   \label{contact}
\end{equation}
Hence, the contact Hamilton's equations \eqref{conham} governed by ${X}_{\bar{E}}$ turn out to be 
\begin{equation}\label{conham+}
\begin{split}
\dot{\mathbf{x}}&=\mathbb{L}E_\mathbf{x}, \\ \dot{\mathbf{x}}^*&= - \nabla_{\mathbf{x}} (\mathbf{x}^* \cdot \mathbb{L}E_\mathbf{x}), \\   \dot{z} &= 0. 
\end{split}
\end{equation}
The time evolution governed by \eqref{conham+} has the following properties: (i) it preserves the contact structure (we shall be more explicit about this in the forthcoming subsection) (ii) the Legendrian submanifold obtained by the first jet section of the Hamiltonian function $E$ defined through  
\begin{equation}\label{subM}
\mathrm{im}(j^1E) =\{(\mathbf{x},\mathbf{x}^*,z)\in \mathbb{M} : \mathbf{x}^*=E_\mathbf{x}, z=E(\mathbf{x})\}
\end{equation}
is invariant, and (iii) the dynamics \eqref{conham+} becomes \eqref{equ1} if restricted to the submanifold $\mathrm{im}(j^1E) $.

\section{Kinetic Dynamics}\label{sec.irr}

 \subsection{From Particle Dynamics to Kinetic Theory}
 
Next comes the task of solving the Hamilton's equation \eqref{equ1} in Poisson picture or, more particularly, the Hamilton's equation (\ref{HamEq-sym}) in the symplectic framework. This is to find trajectories of particles. The collection of all trajectories is called a phase portrait. Our objective is to recognize in the phase portrait generated by \eqref{equ1} a pattern that can be seen as a phase portrait generated by a mesoscopic dynamics that takes into account less details than \eqref{equ1}. The mesoscopic dynamics is represented by
\begin{equation}\label{equ2}
\dot{y}=\mathcal{D}(y)
\end{equation}
where $y$  denotes  the state variable  and $\mathcal{D}$ is an operator \cite{olver86}.

Solutions to (\ref{equ2}) can be compared with results of mesoscopic and macroscopic experimental observations.
How can we pass from \eqref{equ1} to (\ref{equ2}) (in other words, how can we identify $y$ and $\mathcal{D}$)  from an analysis of solutions to \eqref{equ1}? Let us present this.


Following Gibbs \cite{Gibbs}, we begin by putting  Eq.\eqref{equ1} into a form which  could  possibly
be more suitable for recognizing patterns in its phase portrait. The general strategy is to lift \eqref{equ1} to a larger space and search for  patterns in  the  phase portrait generated by the lifted dynamics.

Gibbs lifted \eqref{equ1} into the space of distribution functions. 
In this case, the $y$ variable is the one-particle distribution function $f=f(\mathbf{x},\mathbf{x}^*)$ defined on the cotangent bundle $T^*M$, reversible evolution of which is given by the Vlasov (Liouville) equation, 
\begin{equation}\label{denvlasov}
  \frac{df}{dt} = \{h,f\} = -\nabla_{\mathbf{x}^*} h  \cdot \nabla_\mathbf{x} f + \nabla_{\mathbf{x}} h  \cdot \nabla_{\mathbf{x}^*} f,
\end{equation}
where $\{\bullet,\bullet\}$ is the canonical Poisson bracket \eqref{canPoibra}, and $H$ is the particle Hamiltonian.

The lifted equation  \cite{Liouville}, is again Hamilton's  equation \eqref{equ1}  in the Lie-Poisson picture \cite{mar82,pkg}. The standard physical motivation for making Gibbs' lift is to take statistical view of the particle time evolution and interpret $f$ as a probability distribution function. We can also recall that the original Liouville motivation \cite{Liouville} for writing the Liouville equation was to investigate  relations between ordinary and partial differential equations. The  function  $f$ is not interpreted as a probability distribution function in the Liouville paper. 

There is however also  an alternative physical and mathematical motivation for the Gibbs lift. We see it as a lift bringing  particle Hamilton's mechanics to  a setting with the geometrical structures that have arisen in thermodynamics.

\noindent \textbf{Symplectic Setting.} The Hamiltonian time evolution of particles is a sequence of canonical transformations (transformations of $(\mathbf{x},\mathbf{x}^*)$) that preserve kinematics expressed in the canonical Poisson bracket \eqref{canPoibra}. This promotes the (Lie) group of symplectic automorphisms (diffeomorphisms) on the canonical symplectic manifold $(T^*M,\Omega_M)$ written as  
\begin{equation}
    \text{Diff}_{\text{can}}(T^*M ) =\{\psi \in \text{Diff}(T^*M ):\psi^{\ast }\Omega_M=\Omega_M\}.
\end{equation}
Here, the group multiplication is the composition of diffeomorphisms. The correct Lie algebra of the diffeomorphişsm group $\text{Diff}_{\text{can}}(T^*M )$ is the space of locally Hamiltonian vector fields. By assuming necessary topological constraints, we take the Lie algebra as the space of Hamiltonian vector fields
\begin{equation}
 \mathfrak{X}_{\mathrm{ham}}(T^*M ) =\{X_E \in \mathfrak{X}(T^*M ): \iota_{X_E}\Omega=dE, ~ E \in \mathcal{C}^{\infty}(T^*M ) \}.
\end{equation}
 Here, the Lie bracket $\left[ \bullet,\bullet\right]_{\mathfrak{X}}$, in other words the adjoint representation, is minus the Jacobi-Lie bracket $\left[ \bullet,\bullet\right]_{JL}$ of vector fields:
 \begin{equation}\label{ad-sym}
ad_{X_A}(X_B)=[X_A,X_B]_{\mathfrak{X}}=-[X_A,X_B]_{JL}.
 \end{equation}

Now we define a linear mapping from $\mathcal{C}^{\infty}(T^*M )$  equipped with the canonical Poisson bracket \eqref{poisson-on-ham} to $\mathfrak{X}_{\mathrm{\mathrm{ham}}}(T^*M )$ equipped with the minus of the Jacobi-Lie  bracket:
\begin{equation}\label{epi-onto-Ham}
\varphi:\mathcal{C}^{\infty}(T^*M ) \longrightarrow \mathfrak{X}_{\mathrm{\mathrm{ham}}}(T^*M ), \qquad A \mapsto  X_A.
\end{equation} 
The mapping (\ref{epi-onto-Ham}) is a Lie algebra homomorphism:
\begin{equation}
\left[ \varphi(A),\varphi(B)\right]_{\mathfrak{X}}=-\left[X_{A},X_{B}\right]_{JL} =X_{\left\{A,B\right\} }=\varphi(\left\{ A,B\right\}).
\end{equation}
The mapping $\varphi$ is not injective since the kernel  is the space of constant functions on $T^*M $. Hence, we have 
\begin{equation} \label{X_ham-C}
\mathfrak{X}_{\mathrm{\mathrm{ham}}}(T^*M )\cong \mathcal{C}^{\infty}(T^*M ) / \mathbb{R}.
\end{equation}
Therefore, we obtain two different realizations of the Lie algebra.  
One is the space $\mathfrak{X}_{\mathrm{ham}}(T^*M)$ of Hamiltonian vector fields,  
and the other is the space $\mathcal{C}^{\infty}(T^*M) / \mathbb{R}$ of smooth functions modulo constants.  
These two realizations are related by the Lie algebra homomorphism introduced in \eqref{epi-onto-Ham}.  

The duals of these Lie algebras—namely, the momenta and the densities, respectively—both carry Lie–Poisson brackets, 
and therefore support Lie–Poisson dynamics \eqref{LP-dyn}.  
In particular, the latter case recovers Liouville dynamics precisely in the form of \eqref{denvlasov}.  
Thus, Liouville dynamics admits two equivalent geometric realizations, which we now describe in turn.

\noindent \textbf{Dynamics in Term of Momenta.} Assuming the dimension of the symplectic manifold as $2n$, we fix the symplectic (Liouville) volume $\mathrm{Vol}_\Omega$ defined in \eqref{vol-symp}. Then we consider the dual space $\mathfrak{X}^*_{\mathrm{ham}}(T^*M )$ as one-forms $\Pi$, those making the pairing between the Lie algebra and its dual as
weakly non-degenerate with respect to the $L^{2}$-norm \cite{EsGuRa25,esgu11,esgu12,essu21,gu10}. By dualizing the adjoint action \eqref{ad-sym}, the coadjoint action of the Lie algebra $\mathfrak{g}$ on its dual is computed by 
\begin{equation}
    ad^*: \mathfrak{X}_{\mathrm{ham}}(T^*M )\times \mathfrak{X}^*_{\mathrm{ham}}(T^*M )\longrightarrow \mathfrak{X}^*_{\mathrm{ham}}(T^*M ),\qquad (X_h,\Pi)\mapsto  ad^*_{X_H}\Pi=\mathcal{L}_{X_h}\Pi,
\end{equation}
where $\mathcal{L}_{X_h}$ stands for the Lie derivative operator. 

The (plus) Lie-Poisson bracket \eqref{LP-bra} for the present picture is computed to be
\begin{equation}
\left\{ K ,H \right\}
 \left( \Pi \right) =\int_{T^{\ast }Q }\Pi  \cdot  
\left[ \frac{\delta K}{\delta \Pi },\frac{\delta H}{\delta \Pi  }\right]_{\mathfrak{X}} ~\mathrm{Vol}_\Omega  \label{LP}
\end{equation}
where $\mathrm{Vol}_\Omega$ is the symplectic volume defined in \eqref{vol-symp}. 
Here, $\delta K/\delta \Pi $ and $\delta H/\delta \Pi $
are
regarded to be elements of $\mathfrak{X}_{\mathrm{ham}}(T^*M )$ assuming reflexivity. Then the Lie-Poisson equation is written in terms of the coadjoint action \eqref{LP-dyn} as  
\begin{equation}\label{LP-mom-eq}
\frac{d\Pi}{dt}= - \mathcal{L}_{\delta H/\delta \Pi}\Pi,
\end{equation}
see \cite{gu10,esgu11,esgu12}. 
Considering the representation \eqref{equ2}, we write the Lie-Poisson dynamics \eqref{LP-mom-eq} as
\begin{equation}\label{okort}
\dot{\Pi}=\mathcal{D} (\Pi), \qquad \mathcal{D} =- \mathcal{L}_{\delta H/\delta \Pi}. 
\end{equation}

\noindent \textbf{Dynamics in Term of Densities.} The Lie–Poisson dynamics \eqref{okort} has been studied on the dual space 
$  \mathfrak{X}_{\mathrm{ham}}^*(T^*M)$. Now we write the Lie-Poisson dynamics on the dual space of the Lie algebra $C^\infty(T^*M)/\mathbb{R}$, 
which is naturally identified with the space of densities on $T^*M$. By fixing the symplectic volume, we consider the dual space as $\mathfrak{g}^* \subset C^\infty(T^*M)$. 

The Lie algebra bracket on $ C^\infty(T^*M)/\mathbb{R} $ is the canonical Poisson bracket \eqref{canPoibra} that is
\begin{equation}
    ad: C^\infty(T^*M)/\mathbb{R} \times C^\infty(T^*M)/\mathbb{R} \longrightarrow C^\infty(T^*M)/\mathbb{R} , \qquad (H,F)\mapsto \{H,F\}.  
\end{equation}
Then, due to the computation
\begin{equation}\label{poissoncompo}
\begin{split}
\langle f, ad_{H} F \rangle &= \langle f, \{ H,F\} \rangle \\ &  =\int_{T^*M }f\{ H,F\} ~ d\mu \\ & = 
\int_{T^*M }\{f, H\}~ F ~ d\mu
=\langle ad_{H} ^*f, F \rangle, 
\end{split}
\end{equation}
the coadjoint action of the Lie algebra on its dual space is obtained as 
\begin{equation}
 ad^*: C^\infty(T^*M)/\mathbb{R} \times ( C^\infty(T^*M)/\mathbb{R})^*\longrightarrow ( C^\infty(T^*M)/\mathbb{R})^*, \qquad (H,f)\mapsto ad^*_Hf=\{f,H\}.
\end{equation}

The Lie-Poisson bracket \eqref{LP-bra} on the space $( C^\infty(T^*M)/\mathbb{R})^* \subset C^\infty(T^*M)$ of densities is then obtained as  
\begin{equation} \label{Vlasov-bracket}
\{\mathcal{F},\mathcal{H}\}(f) =\int_{T^*M }f \left\{\frac{\partial \mathcal{F}}{\partial f},\frac{\partial \mathcal{H}}{\partial f}\right\} ~ \mathrm{Vol}_\Omega.
\end{equation}
Hence, the Lie-Poisson dynamics \eqref{LP-mom-eq} for the present picture is computed to be 
\begin{equation}\label{Vlasov-pre}
\frac{d f}{dt}= - ad^*_H f=  \{H,f\}.
\end{equation}
A direct observation gives that, under the Darboux coordinates $(\mathbf{x},\mathbf{x}^*)$ on the cotangent bundle $T^*M$, the Lie-Poisson equation \eqref{Vlasov-pre} is precisely the Liouville equation in \eqref{denvlasov}, see \cite{mar82,pkg}. In terms of the formulation \eqref{equ2}, the Lie-Poisson dynamics \eqref{Vlasov-pre} can be written as 
\begin{equation}
\dot{f}=\mathcal{D} (f), \qquad \mathcal{D} = -ad_{H}^{\ast }(\bullet)=\{H,\bullet\}. 
\end{equation}

As a result, we have two Lie–Poisson brackets, \eqref{LP} and \eqref{Vlasov-bracket}, defined on the space of momenta and the space of densities, respectively,  
as well as two Lie–Poisson dynamics, \eqref{okort} and \eqref{Vlasov-pre}.  
To link these two equivalent realizations of the dynamical equations, we compute the dual of the Lie algebra homomorphism $\varphi$ displayed in \eqref{epi-onto-Ham}. This gives
\begin{equation}\label{epi-onto-Ham+}
\varphi^*:\mathfrak{X}^*_{\mathrm{ham}}(T^*M ) \longrightarrow \mathcal{C}^{\infty}(T^*M ), \qquad \Pi \mapsto f =\mathrm{div}\,\Omega_M^\sharp(\Pi),
\end{equation}
where $\Omega_M^\sharp$ is the musical isomorphism induced by the canonical symplectic two-form, and  
$\mathrm{div}$ is the divergence of the vector field $\Omega_M^\sharp(\Pi)$ with respect to the symplectic volume. Being the dual of a Lie algebra homomorphism, the mapping $\varphi^*$ is itself a Poisson map,  
and therefore respects both the Lie–Poisson structures and the corresponding Lie–Poisson dynamics.

In terms of the Darboux coordinates $(\mathbf{x},\mathbf{x}^*)$, we write a one-form as 
\begin{equation}\label{one-form}
\Pi =\Pi_{\mathbf{x}}  \cdot
d\mathbf{x}+ \Pi
_{\mathbf{x}^*}  \cdot d\mathbf{x}^*, 
\end{equation}
hence the associated density is computed to be  
\begin{equation}
f(\mathbf{x},\mathbf{x}^*)= \mathrm{div}\Omega_M^\sharp(\Pi) =\nabla
_{\mathbf{x}^*}\cdot \Pi_{\mathbf{x}} -\nabla _{\mathbf{x}}\cdot \Pi_{\mathbf{x}^*} . \label{density}
\end{equation}
This computation also characterizes how the dual space $\mathfrak{g}^*$ is realized within the space $C^\infty(T^*M)$ of smooth functions.

\noindent \textbf{Choosing Contact Particle Dynamics.}
In the sequel, we shall approach the geometric and Hamiltonian analysis of the Liouville equation in a more general setting.  
Instead of considering symplectic dynamics (presented in Subsection \ref{sec.rev}) at the particle level, we shall begin with contact dynamics (presented in Subsection \ref{sec:contact-dyn}) at the particle level and then lift it to kinetic dynamics.  
We refer to this framework as a new class of contact kinetic models.  

The motivation behind choosing the contact framework comes from~\cite{VanHove} and the footnote in~\cite{marsden1982hamiltonian}, where the correct configuration space of the plasma was interpreted as the group of contact diffeomorphisms (contactomorphisms) on $T^*Q \times \mathbb{R}$.  
For existing formulations of contact kinetic dynamics, we refer the reader to~\cite{AtEsLeSa,EsGeGrGuPaSu,esgu11}.  

The kinetic equations presented in this paper extend those in the literature by introducing the phase-space volume as an independent variable, which may provide a geometric mechanism for entropy in contact kinetic dynamics.  
As we shall point out, this extension is achieved by treating the group of contact diffeomorphisms as a semi-direct product consisting of pairs of contact diffeomorphisms and a conformal factor.  
This structural ingredient is absent from existing formulations.  

In particular, we shall demonstrate how the present work not only generalizes the available geometric formulations of contact kinetic dynamics, but also recovers the classical Liouville (Vlasov) dynamics as a particular instance.

\subsection{Contact Group, Contact Algebra, and the Dual Space}
\label{sec:gr}

We first determine the group of contact diffeomorphisms on a contact manifold, then derive its Lie algebra, and finally formulate the corresponding Lie–Poisson dynamics on the dual space.

\noindent \textbf{Group Structure.} Consider a contact manifold $(\mathbb{M},\eta)$. A  diffeomorphism (that is a differentiable automorphism) $\varphi$ on $(\mathbb{M},\eta)$ is called  contact if it preserves the contact structure, that is
\begin{equation}
\varphi ^{\ast }\eta =\gamma_\varphi \eta,
\end{equation} 
for some real-valued function $\gamma_\varphi$. Hence, we denote a contact diffeomorphism (also called contactomorphism) by a two-tuple $(\varphi,\gamma_\varphi)$. The
group of contact diffeomorphisms on $\mathbb{M}$ is 
\begin{equation}\label{Diff-con}
G={\rm Diff}_{\mathrm{con}} (\mathbb{M} ) =\left\{ (\varphi,\gamma_\varphi) \in {\rm Diff} ( \mathbb{M})\times   \mathcal{F}
( \mathbb{M} ):\varphi ^{\ast }\eta =\gamma_\varphi  \eta  \right\} .
\end{equation} 
To have the group operation on ${\rm Diff}_{\mathrm{con}} (\mathbb{M} )$, we take two elements $(\varphi,\gamma_\varphi)$ and $(\varrho,\gamma_\varrho)$ from the group and perform the following computation:
\begin{equation}
\begin{split}
(\varrho \circ \varphi)^*(\eta) & = \varphi^* \varrho ^* (\eta) = \varphi^* (\gamma_\varrho \eta) \\
& = \varphi^*(\gamma_\varrho) \eta + \gamma_\varrho \varphi^*(\eta)
\\
& = \varphi^*(\gamma_\varrho) \eta + \gamma_\varrho \gamma_\varphi \eta .
\end{split}
\end{equation} 
This calculation gives the semi-direct product Lie group structure on $G={\rm Diff}_{\mathrm{con}} (\mathbb{M} )$ as
\begin{equation}
G \times G\longrightarrow G,\qquad (\varrho,\gamma_\varrho) \cdot (\varphi,\gamma_\varphi) = \big(\varrho \circ \varphi, \varphi^*(\gamma_\varrho) + \gamma_\varrho \gamma_\varphi \big ).
\end{equation} 


\noindent \textbf{Lie Algebra.} Consider a contact manifold $( \mathbb{M},\eta) $ and the group of contact transformations $G={\rm Diff}_{\mathrm{con}} (\mathbb{M} ) $ in \eqref{Diff-con}. The Lie algebra of $G$ is the space of infinitesimal contact transformations: 
\begin{equation}\label{algcon}
\mathfrak{g}=\mathfrak{X}_{\mathrm{con}} ( \mathbb{M}  ) =\left\{ (X,\lambda_X)\in \mathfrak{X} ( 
\mathbb{M} )\times  \mathcal{F} ( \mathbb{M}  ):\mathcal{L}_{X}\eta  =-\lambda_X \eta  \right\} .  
\end{equation}
Notice that we denote an element $(X,\lambda_X)$ of the Lie algebra by a pair consisting of a vector field $X$ and a real-valued function $\lambda_X $ called conformal factor. To have the Lie algebra bracket on $\mathfrak{g}=\mathfrak{X}_{\mathrm{con}}(\mathbb{M})$, we consider two elements $(X,\lambda_X)$ and $(Y,\lambda_Y)$ and calculate  
\begin{equation} 
\begin{split}
\mathcal{L}_{[X,Y]} \eta & = \mathcal{L}_X \mathcal{L}_Y \eta - \mathcal{L}_Y \mathcal{L}_X \eta     
\\
& = -  \mathcal{L}_X (\lambda_Y \eta) +  \mathcal{L}_Y (\lambda_X \eta)
\\
& = - \mathcal{L}_X (\lambda_Y)  \eta  - \lambda_Y \mathcal{L}_X ( \eta) +  \mathcal{L}_Y (\lambda_X)  \eta +  \lambda_X  \mathcal{L}_Y (\eta)
\\
& = -
\mathcal{L}_X (\lambda_Y)  \eta  + \lambda_Y \lambda_X  \eta  +  \mathcal{L}_Y (\lambda_X)  \eta 
- \lambda_X \lambda_Y\eta 
\\
& = \big(\mathcal{L}_Y (\lambda_X) - \mathcal{L}_X (\lambda_Y)\big)\eta .
\end{split}
\end{equation}
We assume the Lie bracket of vector fields in $\mathfrak{X}_{\mathrm{con}} ( \mathbb{M}  )$ as the minus of the Jacobi-Lie bracket. Hence, we arrive at the semi-direct product Lie algebra structure on $\mathfrak{g}=\mathfrak{X}_{\mathrm{con}} ( \mathbb{M}  )$ as 
\begin{equation}\label{algebra-new}
\mathfrak{g}\times \mathfrak{g}\longrightarrow \mathfrak{g},\qquad 
[(X,\lambda_X),(Y,\lambda_Y)]_{\mathfrak{g}}  = \big(-[X,Y], \mathcal{L}_Y (\lambda_X) - \mathcal{L}_X (\lambda_Y) \big).
\end{equation} 

\noindent \textbf{The Dual Space.} To formulate the dual picture, we first fix the contact volume $\mathrm{Vol}_\eta$ defined in \eqref{vol-cont}. Then we determine the dual space of the infinitesimal contact transformations $\mathfrak{X}_{\mathrm{con}} ( \mathbb{M}  )$ as a subspace of the product space:
\begin{equation}
    \mathfrak{g}^*=\mathfrak{X}^*_{\mathrm{con}} ( \mathbb{M}  ) \subset \Gamma^1(\mathbb{M})\otimes C^\infty(\mathbb{M})
\end{equation}
of one-form sections $\Gamma^1(\mathbb{M})$ and smooth functions $C^\infty(\mathbb{M})$ making the pairing
\begin{equation} \label{pairing}
\langle (\Pi,S),(X,\lambda_X) \rangle = \int_{\mathbb{M} } \Pi\cdot X ~ \mathrm{Vol}_\eta +  \int_\mathbb{M}  S  \lambda_X ~ \mathrm{Vol}_\eta
\end{equation}
weakly non-degenerate. 

\noindent \textbf{Coadjoint Action.} According to \eqref{coad}, the coadjoint action of the Lie algebra $ \mathfrak{g}=\mathfrak{X}_{\mathrm{con}} ( \mathbb{M}  )$ on the dual space $ \mathfrak{g}^*=\mathfrak{X}^*_{\mathrm{con}} ( \mathbb{M}  )$ is defined as follows
\begin{equation} 
ad^*:\mathfrak{g}\times \mathfrak{g}^*  \longrightarrow\mathfrak{g}^*, \quad 
  \big \langle ad^{\ast }_{(X,\lambda_X)} (\Pi,S), (Y,\lambda_Y) \big\rangle = 
 \big\langle \Pi , ad_{(X,\lambda_X)} (Y,\lambda_Y) \big\rangle. 
\end{equation}
In order to compute the coadjoint action in explicit terms, we do the following computation: 
\begin{equation}
\begin{split}
&\left\langle ad_{(X,\lambda_X)} ^* (\Pi,S)  , (Y,\lambda_Y) \right\rangle = \left\langle  (\Pi,S) , ad_{(X,\lambda_X)} (Y,\lambda_Y)  \right\rangle \\ 
&\qquad = \big \langle  (\Pi,S) ,  \big(-[X,Y], \mathcal{L}_Y (\lambda_X) - \mathcal{L}_X (\lambda_Y) \big) \big  \rangle
\\ 
&\qquad = - \int \langle \Pi,[X,Y] \rangle \mathrm{Vol}_\eta + 
\int S \mathcal{L}_Y (\lambda_X)  \mathrm{Vol}_\eta 
\\ 
&\qquad \qquad \qquad -
\int S \mathcal{L}_X (\lambda_Y)   \mathrm{Vol}_\eta 
\\ 
&\qquad = \int \langle \mathcal{L}_X  \Pi +\mathrm{div}(X)  \Pi , Y \rangle  \mathrm{Vol}_\eta + 
\int S \iota_Y (d\lambda_X)  \mathrm{Vol}_\eta \\ 
&\qquad \qquad \qquad - 
\int S \iota_X (d\lambda_Y)   \mathrm{Vol}_\eta
\\ 
&\qquad = \int \langle \mathcal{L}_X  \Pi +\mathrm{div}(X)  \Pi , Y \rangle  \mathrm{Vol}_\eta +
\int \langle S  d\lambda_X , Y \rangle \mathrm{Vol}_\eta \\ 
&\qquad \qquad \qquad - 
\int S d\lambda_Y \wedge   \iota_X \mathrm{Vol}_\eta
\\ 
&\qquad = \int \langle \mathcal{L}_X  \Pi +\mathrm{div}(X)  \Pi , Y \rangle  \mathrm{Vol}_\eta  +
\int \langle S  d\lambda_X , Y \rangle \mathrm{Vol}_\eta  \\
&\qquad \qquad 
+
\int  \lambda_Y dS\wedge   \iota_X \mathrm{Vol}_\eta +
\int  \lambda_Y S   d\iota_X \mathrm{Vol}_\eta 
\\ 
&\qquad = \int \langle \mathcal{L}_X  \Pi +\mathrm{div}(X)  \Pi +S  d\lambda_X  , Y \rangle  \mathrm{Vol}_\eta 
+
\int  \lambda_Y \iota_XdS  \mathrm{Vol}_\eta \\ 
&\qquad \qquad \qquad +
\int  \lambda_Y S   \mathcal{L}_X \mathrm{Vol}_\eta 
\\ 
&\qquad = \int \langle \mathcal{L}_X  \Pi +\mathrm{div}(X)  \Pi - \lambda_X dS, Y \rangle  \mathrm{Vol}_\eta  
+
\int  \lambda_Y \iota_XdS  \mathrm{Vol}_\eta \\ 
&\qquad \qquad \qquad +
\int  \lambda_Y S   \mathrm{div}(X) \mathrm{Vol}_\eta 
\\  
&\qquad = \int \langle \mathcal{L}_X  \Pi +\mathrm{div}(X)  \Pi - \lambda_X dS, Y \rangle  \mathrm{Vol}_\eta  
\\ 
&\qquad \qquad \qquad  +
\int  \lambda_Y \big( \iota_XdS + S   \mathrm{div}(X) \big)  \mathrm{Vol}_\eta, 
\\ 
\end{split}
\end{equation} 
which means that the coadjoint action acts as
\begin{equation}\label{coad-contacta}
    ad_{(X,\lambda_X)} ^* (\Pi,S)= \big(\mathcal{L}_X  \Pi +\mathrm{div}(X)  \Pi - \lambda_X dS, X(S) + S   \mathrm{div}(X)).
\end{equation}

\noindent \textbf{Lie-Poisson Bracket.} Being the dual space of a Lie algebra, $\mathfrak{X}_{\mathrm{con}}^*(\mathbb{M})$ admits Lie-Poisson bracket \cite{holm11,holm2009geometric,MaRa99}. More precisely, referring to \eqref{LP-bra}, given two functionals $A$ and $B$ on $\mathfrak{g}^*=\mathfrak{X}_{\mathrm{con}}^*(\mathbb{M})$, the Lie-Poisson bracket on $\mathfrak{g}^*=\mathfrak{X}_{\mathrm{con}}^*(\mathbb{M})$ is
\begin{equation}\label{LPBr}
\begin{split}
\left\{ A,B\right\} ^{\mathfrak{g}^*} (\Pi,S) & = \int  \Big\langle (\Pi,S) , ad_{\delta A / \delta (\Pi,S)} 
 \frac{\delta B}{\delta (\Pi,S)} \Big\rangle \mathrm{Vol}_\eta \\& =  \int \Big\langle (\Pi ,S), \big[\big(\frac{\delta A}{\delta \Pi} , \frac{\delta A}{\delta S} \big), 
 \big(\frac{\delta B}{\delta \Pi} , \frac{\delta B}{\delta S} \big)  \big ] \Big\rangle \mathrm{Vol}_\eta
\\& =   - \int \Big\langle (\Pi , S), \big[\big(\frac{\delta B}{\delta \Pi} , \frac{\delta B}{\delta S} \big), 
 \big(\frac{\delta A}{\delta \Pi} , \frac{\delta A}{\delta S} \big)  \big ] \Big\rangle \mathrm{Vol}_\eta\\& 
 =   - \int \Big\langle ad^*_{(\delta B/\delta \Pi, \delta B/\delta S)}  (\Pi , S),  
 \big(\frac{\delta A}{\delta \Pi} , \frac{\delta A}{\delta S} \big)   \Big\rangle \mathrm{Vol}_\eta
\end{split}
\end{equation}%
where  ${\delta A}/{\delta \Pi}$ stands for the Fr\'echet derivative of the functional $A$ with respect to $\Pi$ whereas $\delta A/\delta S$ stands for the Fr\'echet derivative of the functional $A$ with respect to $S$.  Notice  that we have also assumed the reflexivity and the following identification  
\begin{equation}
     \frac{\delta A}{\delta (\Pi,S)}=\left(\frac{\delta A}{\delta \Pi} , \frac{\delta A}{\delta S} \right)\in \mathfrak{X}_{\mathrm{con-ham}} (\mathbb{M}).
\end{equation}

\noindent \textbf{A New Contact Kinetic Dynamics.}
Given a Hamiltonian functional $\mathcal{H}$, the Lie-Poisson dynamics is governed by the Lie-Poisson equations computed in terms of the coadjoint action, that is,
\begin{equation}
\begin{split}    
\frac{d}{dt} (\Pi,S) &= \{(\Pi,S), \mathcal{H}\}^{\mathfrak{g}^*} =- ad_{\delta \mathcal{H} / \delta (\Pi,S) }^{\ast }(\Pi,S)
\\&=- ad^*_{\big(\frac{\delta \mathcal{H}}{\delta \Pi}, \frac{\delta \mathcal{H}}{\delta S}\big)}(\Pi,S)
\\&=- \big(\mathcal{L}_{\frac{\delta \mathcal{H}}{\delta \Pi}}  \Pi +\mathrm{div}(\frac{\delta \mathcal{H}}{\delta \Pi})  \Pi - \frac{\delta \mathcal{H}}{\delta S} dS, \frac{\delta \mathcal{H}}{\delta \Pi}(S) + S   \mathrm{div}(\frac{\delta \mathcal{H}}{\delta \Pi})).
\end{split}
\end{equation}
Therefore, the Lie-Poisson equations for the kinetic dynamics of a bunch of particles moving through infinitesimal contact transformations are 
\begin{equation}\label{LPEq-}
\begin{split}    
\frac{d\Pi}{dt}  & = -  \mathcal{L}_{\frac{\delta \mathcal{H}}{\delta \Pi}}  \Pi +\mathrm{div}(\frac{\delta \mathcal{H}}{\delta \Pi})  \Pi - \frac{\delta \mathcal{H}}{\delta S} dS,\\
\frac{dS}{dt}  &= - \frac{\delta \mathcal{H}}{\delta \Pi}(S) + S   \mathrm{div}(\frac{\delta \mathcal{H}}{\delta \Pi}).
\end{split}
\end{equation}

Assume a Hamiltonian functional defined by means of the infinitesimal contact transformation $(X,\lambda_X)$ as 
 \begin{equation} \label{Ham-a}
 \mathcal{H}(\Pi,S)=\int \langle \Pi,X \rangle \mathrm{Vol}_\eta + \int   S \lambda_X   \mathrm{Vol}_\eta.
\end{equation}
Here, the Fr\'{e}chet derivative $\delta \mathcal{H}/ \delta \Pi$ of $\mathcal{H}$ with respect to the momenta becomes the vector field $X$ while the Fr\'{e}chet derivative $\delta\mathcal{H}/ \delta S$ of $\mathcal{H}$ with respect to scalar $S$ is $\lambda_X$. In this case, the Lie-Poisson equation \eqref{LPEq-} takes the form of
\begin{equation}\label{LPEq-+}
\begin{split}    
\frac{d\Pi}{dt}  & = -  \mathcal{L}_{X}  \Pi +\mathrm{div}(X)  \Pi - \lambda_X dS,\\
\frac{dS}{dt}  &= - X(S) + S   \mathrm{div}(X).
\end{split}
\end{equation}

The novel kinetic formulations \eqref{LPEq-} and \eqref{LPEq-+} differ from the existing contact kinetic formulations in the literature~\cite{esgu11,esgu12,EsGeGrGuPaSu} by incorporating the dynamics of the additional variable $S$. To have the contact kinetic dynamics those available in the literature, one needs to take $S$ constant, then the equations reduce to  
\begin{equation}\label{LPEq-++} 
\frac{d}{dt} \Pi = -  \mathcal{L}_{X} \Pi + \mathrm{div}(X)\,\Pi .
\end{equation} 
Moreover, if one further assumes that the dynamics is generated by divergence-free vector fields, this reduces to  
\begin{equation}\label{LPEq-++--} 
\frac{d}{dt} \Pi = - \mathcal{L}_{X} \Pi , 
\end{equation}
which is precisely the Liouville dynamics in momentum formulation, as displayed in~\eqref{LP-mom-eq}.

To investigate how $S$ evolves in the contact kinetic formulations \eqref{LPEq-} and \eqref{LPEq-+}, we compute the following integral: 
\begin{equation}\label{hope-new}
\begin{split}
    \frac{d}{dt}\int S \mathrm{Vol}_\eta &= - \int_\Omega X(S) \mathrm{Vol}_\eta+ \int_\Omega S   \mathrm{div}(X) \mathrm{Vol}_\eta \\ &= - \int_{\delta \Omega} S X \cdot \mathbf{n} dA +\int_\Omega S   \mathrm{div}(X) \mathrm{Vol}_\eta \\ &= \int_\Omega S   \mathrm{div}(X) \mathrm{Vol}_\eta,
\end{split}
\end{equation}
where we have assumed that $X$ is parallel to the boundaries. This computation gives that if $\mathrm{div}(X)$ is positive (negative) then $S$ is increasing (resp. decreasing). If $\mathrm{div}(X)=0$, as in the case of usual symplectic mechanics, then $S$ turns out to be constant. $S$ thus corresponds to the phase-space volume occupied by the particles. As changes in the phase-space volume are related to the entropy production \cite{netocny2002}, we conclude that $S$ can be used as a measure of entropy production. Moreover, for the contact Hamiltonian vector field \eqref{conham}, the divergence is proportional to the derivative of the Hamiltonian function with respect to the $z$-coordinate (or action of the Reeb vector field on the Hamiltonian function.)

 \subsection{Hamiltonian Contact Kinetic Dynamics: Momenta Formulation} \label{sec:Ham-mom}

In the previous subsection, we considered the group 
$G={\rm Diff}_{\mathrm{con}}(\mathbb{M})$ of contact diffeomorphisms introduced in~\eqref{Diff-con}.  
Its Lie algebra is given by the space  
$\mathfrak{g}=\mathfrak{X}_{\mathrm{con}}(\mathbb{M})$ of infinitesimal contact transformations, as in~\eqref{algcon}.  

In the present section, we restrict our attention to contact Hamiltonian vector fields.  
That is, we regard the space of contact Hamiltonian vector fields as a Lie subalgebra of  
$\mathfrak{g}=\mathfrak{X}_{\mathrm{con}}(\mathbb{M})$.  
This allows us to determine the corresponding Lie–Poisson formulation on the dual space.  
The main advantage of focusing on contact Hamiltonian vector fields is that it provides  
a natural and geometrically consistent framework to express kinetic dynamics directly  
in terms of density functions.



\noindent \textbf{Lie Algebra of Contact Hamiltonian Vector Fields.} Consider a contact manifold $(\mathbb{M},\eta)$, and a contact Hamiltonian vector field $X_{H}$ defined by~\eqref{con-Ham-v-f}.  
A direct computation shows that the contact one-form changes by a conformal factor under the action of $X_{H}$:
\begin{equation}\label{L-X-eta}
\mathcal{L}_{X_{H}}\eta
= d(\iota_{X_{H}}\eta)+\iota_{X_{H}}d\eta
= -\mathcal{R}(H)\,\eta.
\end{equation}
Hence, the pair $(X_{H},\mathcal{R}(H))$ constitutes an infinitesimal contact transformation,  
and thus belongs to the Lie algebra $\mathfrak{g}=\mathfrak{X}_{\mathrm{con}}(\mathbb{M})$ in~\eqref{algcon}.

Let us examine the algebraic character of Hamiltonian vector fields within the Lie algebra structure.  
For two Hamiltonian vector fields determined by the pairs $\big(X_{F},\mathcal{R}(F)\big)$ and $\big(X_{H},\mathcal{R}(H)\big)$,  
the Lie bracket~\eqref{algebra-new} reads
\begin{equation} \label{comco}
\begin{split}
&\Big[\big(X_{F},\mathcal{R}(F)\big),\big(X_{H},\mathcal{R}(H)\big)\Big]  \\
& \qquad = \big(-[X_{F},X_{H}], \mathcal{L}_{X_{H}} (\mathcal{R}(F)) - \mathcal{L}_{X_{F}} (\mathcal{R}(H)) \big) \\
& \qquad  = \big( X_{\{F,H\}^{(C)}}, \{\mathcal{R}(F),H\}^{(C)}-\mathcal{R}(F)\mathcal{R}(H) -\{\mathcal{R}(H),F\}^{(C)} \\
& \qquad\qquad+\mathcal{R}(H)\mathcal{R}(F)\big) \\
& \qquad  = \big( X_{\{F,H\}^{(C)}}, \{\mathcal{R}(F),H\}^{(C)} + \{F,\mathcal{R}(H)\}^{(C)} \big) \\
& \qquad  = \big( X_{\{F,H\}^{(C)}}, \mathcal{R}\{F,H\}^{(C)} \big).
\end{split}
\end{equation}
Here, in passing to the second line we applied the identities \eqref{pelin} and \eqref{pelin2},  
while the simplification to the fourth line follows from the distributivity property of the Reeb vector field:
\begin{equation}
 \mathcal{R}\{F,H\}^{(C)} = \{\mathcal{R}(F),H\}^{(C)} + \{F,\mathcal{R}(H)\}^{(C)}.
\end{equation}
Thus, the space of contact Hamiltonian vector fields
\begin{equation}
\begin{split}
\mathfrak{g}_{\mathrm{ham}}
&=\mathfrak{X}_{\mathrm{con-ham}}(\mathbb{M}) \\
&= \Big\{ (X_H,\mathcal{R}(H)) \in \mathfrak{g} \; : \; 
\iota_{X_{H}}\eta =-H,\quad  
\iota_{X_{H}}d\eta =dH-\mathcal{R}(H)\eta \Big\}
\end{split}
\end{equation}
is a Lie subalgebra of the space $\mathfrak{g}=\mathfrak{X}_{\mathrm{con}}(\mathbb{M})$ of infinitesimal contact transformations. 
Another consequence of the calculation in \eqref{comco} is that we have the following map
\begin{equation}\label{vartheta}
    \vartheta: C^\infty (\mathbb{M})\longrightarrow \mathfrak{g}_{\mathrm{ham}}= \mathfrak{X}_{\mathrm{con-ham}} (\mathbb{M}),\qquad H \mapsto (X_{H},\mathcal{R}(H)).
\end{equation}
The kernel of the mapping $\vartheta$ is trivial; therefore, it defines a Lie algebra isomorphism.
Recall that, in the symplectic case, we analogously have the mapping $\varphi$ introduced in \eqref{epi-onto-Ham},
whose kernel, however, is non-trivial. 

\noindent \textbf{The Dual Space.}
Recall the pairing \eqref{pairing} between the infinitesimal contact transformation space $\mathfrak{g}=\mathfrak{X}_{\mathrm{con}} (\mathbb{M})$ and its dual space $\mathfrak{g}^*=\mathfrak{X}^*_{\mathrm{con}}(\mathbb{M})$. Let us examine this for the space $\mathfrak{g}_{\mathrm{ham}}=\mathfrak{X}_{\mathrm{con-ham}} (\mathbb{M})$  of contact Hamiltonian vector fields. In this case, we have
\begin{equation}\label{peper}
 \left\langle (\Pi, S), (X_{H},\mathcal{R}(H)) \right\rangle  = \int_{\mathbb{M}} \langle \Pi, X_{H} 
 \rangle   \mathrm{Vol}_\eta +  \int_{\mathbb{M}} S  \mathcal{R}(H)  \mathrm{Vol}_\eta .
 \end{equation}
For the first term on the right-hand side of \eqref{peper}, we compute 
 \begin{equation}\label{dual-p}
 \begin{split}
  \int_{\mathbb{M}} \langle &\Pi, X_{H} 
 \rangle   \mathrm{Vol}_\eta =\int_{\mathbb{M}} \langle \Pi, \sharp(dH) - (\mathcal{R}(H)+H )\mathcal{R}
 \rangle  \mathrm{Vol}_\eta 
\\
&= - \int_{\mathbb{M}} \langle \sharp\Pi, dH  \rangle  \mathrm{Vol}_\eta  + \int_{\mathbb{M}} (\mathcal{R}(H)+H ) \langle \sharp\Pi,\eta  \rangle  \mathrm{Vol}_\eta  
\\
&= - \int_{\mathbb{M}} \big(\iota_{\sharp\Pi}  dH \big)  \mathrm{Vol}_\eta  + \int_{\mathbb{M}}  \mathcal{R}(H)  \langle \sharp\Pi,\eta  \rangle  \mathrm{Vol}_\eta  + \int_{\mathbb{M}}   H   \langle \sharp\Pi,\eta  \rangle  \mathrm{Vol}_\eta 
 \\&=  -
 \int_{\mathbb{M}}  dH \iota_{\sharp\Pi} \mathrm{Vol}_\eta + 
 \int_{\mathbb{M}}  \iota_{\mathcal{R}}dH   \langle \sharp\Pi,\eta  \rangle  \mathrm{Vol}_\eta + \int_{\mathbb{M}}   H   \langle \sharp\Pi,\eta  \rangle  \mathrm{Vol}_\eta 
  \\&=  \int_{\mathbb{M}}  H d \iota_{\sharp\Pi} \mathrm{Vol}_\eta
  + \int_{\mathbb{M}} \langle \sharp\Pi,\eta  \rangle dH \wedge  \iota_{\mathcal{R}} \mathrm{Vol}_\eta + \int_{\mathbb{M}}   H   \langle \sharp\Pi,\eta  \rangle  \mathrm{Vol}_\eta 
   \\&= \int_{\mathbb{M}}  H d \iota_{\sharp\Pi} \mathrm{Vol}_\eta
  - \int_{\mathbb{M}} H d\langle \sharp\Pi,\eta  \rangle \wedge  \iota_{\mathcal{R}} \mathrm{Vol}_\eta 
  -\int_{\mathbb{M}} H  \langle \sharp\Pi,\eta  \rangle    d\iota_{\mathcal{R}} \mathrm{Vol}_\eta
  \\& \qquad + \int_{\mathbb{M}}   H   \langle \sharp\Pi,\eta  \rangle  \mathrm{Vol}_\eta   
  \\ &= 
  \int_{\mathbb{M}}  H \Big( \mathrm{div}(\sharp\Pi) \mathrm{Vol}_\eta
  -   \iota_{\mathcal{R}}d\langle \sharp\Pi,\eta  \rangle  - \langle \sharp\Pi,\eta  \rangle    \mathrm{div}(\mathcal{R})    + \langle \sharp\Pi,\eta  \rangle \Big)  \mathrm{Vol}_\eta    
 \\ &= 
  \int_{\mathbb{M}}  H \Big( \mathrm{div}(\sharp\Pi) 
  -   \mathcal{L}_\mathcal{R} \langle \sharp\Pi,\eta  \rangle       + \langle \sharp\Pi,\eta  \rangle \Big)  \mathrm{Vol}_\eta. 
\end{split}
\end{equation}
Here, in the first line, we have employed \eqref{3-def2}. Note that $\mathrm{div}$ stands for the divergence with respect to the contact volume $\mathrm{Vol}_\eta$. 
We compute the second term on the right-hand side of \eqref{peper} as 
\begin{equation}\label{dual-p2} 
\begin{split}
 \int_{\mathbb{M}} S  \mathcal{R}(H)  \mathrm{Vol}_\eta & = \int_{\mathbb{M}} S  \iota_ \mathcal{R}(dH)  \mathrm{Vol}_\eta = \int_{\mathbb{M}} S  ~ dH  \wedge \iota_ \mathcal{R} \mathrm{Vol}_\eta \\ & = - \int_{\mathbb{M}} H  dS  \wedge \iota_ \mathcal{R} \mathrm{Vol}_\eta - \int_{\mathbb{M}} S  ~ H   d\iota_ \mathcal{R} \mathrm{Vol}_\eta 
 \\ & =  - \int_{\mathbb{M}} H \iota_ \mathcal{R} dS    \mathrm{Vol}_\eta - \int_{\mathbb{M}} S   H   \mathcal{L}_ \mathcal{R} \mathrm{Vol}_\eta 
  \\ & = - \int_{\mathbb{M}}  H \big(\iota_ \mathcal{R} dS + S\mathrm{div}(\mathcal{R}) \big) \mathrm{Vol}_\eta
   \\ & = - \int_{\mathbb{M}}  H  \iota_ \mathcal{R} dS \mathrm{Vol}_\eta. 
 \end{split}
 \end{equation}
Here, the second term in the fourth line is zero since the divergence $\mathrm{div}(\mathcal{R})$ of the Reeb vector field is identically zero as it can be easily verified from \eqref{Reeb}.  
The calculations in \eqref{dual-p} and \eqref{dual-p2} yield the pairing \eqref{peper} as 
\begin{equation}\label{dual-p1.collected} 
\begin{split}
& \left\langle (\Pi, S), (X_{H},\mathcal{R}(H))  \right\rangle \\ & \qquad =  \int_{\mathbb{M}}  H \Big( \mathrm{div}(\sharp\Pi) 
  -   \mathcal{L}_\mathcal{R} \langle \sharp\Pi,\eta  \rangle       + \langle \sharp\Pi,\eta  \rangle -\iota_ \mathcal{R} dS  \Big)  \mathrm{Vol}_\eta,
  \end{split}
 \end{equation}
and the pairing is non-degenerate if 
\begin{equation}
   \mathrm{div}(\sharp\Pi) 
  -   \mathcal{L}_\mathcal{R} \langle \sharp\Pi,\eta  \rangle       + \langle \sharp\Pi,\eta  \rangle -\iota_ \mathcal{R} dS   \neq 0.  
\end{equation}

\noindent \textbf{The Dual Mapping.} To pass the density formulation of the contact kinetic dynamics. We recall the Lie algebra isomorphism $\vartheta$ in \eqref{vartheta} and write its dual map as 
\begin{equation}
\vartheta^*: \mathfrak{g}^*_{\mathrm{ham}} (\mathbb{M}) \longrightarrow \mathrm{Den} (\mathbb{M}),\qquad (\Pi,S) \mapsto f \mathrm{Vol}_\eta. 
\end{equation}
Being the dual of a Lie algebra isomorphism, the dual mapping is a momentum and a Poisson mapping.
Fixing the contact volume, the calculation \eqref{dual-p1.collected} reads the density function $f$ as
\begin{equation}\label{Psi-*}
f=\vartheta^*(\Pi,S)=\mathrm{div}(\sharp\Pi) 
  -   \mathcal{L}_\mathcal{R} \langle \sharp\Pi,\eta  \rangle       + \langle \sharp\Pi,\eta  \rangle -\iota_ \mathcal{R} dS.
\end{equation}

In terms of the Darboux coordinates $(\mathbf{x},\mathbf{x}^*,z)$ on contact manifold $\mathbb{M}$, we can compute the density function $f$ from \eqref{Psi-*} as follows. Consider a one-form section 
\begin{equation}
\Pi  =\Pi_{\mathbf{x}} \cdot  d\mathbf{x} + \Pi_{\mathbf{x}^*}\cdot d\mathbf{x}^* + \Pi_z dz.
\end{equation}
Then, using \eqref{loc-flat-sharp}, we compute
\begin{equation}
\begin{split}
\langle \sharp \Pi,\eta  \rangle = & \Big \langle \Pi_{\mathbf{x}^*}\cdot \nabla_\mathbf{x} - \big(\Pi_{\mathbf{x}}+ \Pi_z \mathbf{x}^* \big)\cdot \nabla_{\mathbf{x}^*} \\ & \qquad  + \big( \Pi_z+ \Pi_{\mathbf{x}^*}\cdot \mathbf{x}^*\big)\nabla_z,  dz- \mathbf{x}^* \cdot d \mathbf{x}\Big  \rangle \\
= &~ \Pi_z+ \Pi_{\mathbf{x}^*}\cdot \mathbf{x}^* - \mathbf{x}^* \cdot \Pi_{\mathbf{x}^*} =\Pi_z, 
\end{split}
\end{equation}
hence the second Lie-derivative term on the right-hand side of \eqref{Psi-*} is 
\begin{equation}
  \mathcal{L}_\mathcal{R} \langle \sharp\Pi,\eta  \rangle    = \mathcal{L}_\mathcal{R} \Pi_z  =\iota_{\mathcal{R}}d\Pi_z=\nabla_z \Pi_z. 
\end{equation}
The contact divergence of $\Pi $ is
\begin{equation}
 \mathrm{div}(\sharp\Pi)  =\nabla_{\mathbf{x}} \cdot \Pi_{\mathbf{x}^*} - \nabla_{\mathbf{x}^*} \cdot \Pi_{\mathbf{x}}  -n\Pi_z -\mathbf{x}^*\cdot \nabla_{\mathbf{x}^*}   + \nabla_z \Pi_z +\nabla_z (\mathbf{x}^*\cdot \Pi_{\mathbf{x}^*}),
\end{equation}
while the fourth term is $\iota_ \mathcal{R} dS=\mathcal{R}(S)=\nabla_z S$.  
Adding all of these terms, we arrive at the definition of the density function
\begin{equation}\label{density-f}
    f=  \nabla_{\mathbf{x}} \cdot \Pi_{\mathbf{x}^*} - \nabla_{\mathbf{x}^*} \cdot \Pi_{\mathbf{x}^*}   - \mathbf{x}^* \cdot \nabla_{\mathbf{x}^*} \Pi_z + \nabla_{z}(\mathbf{x}^*\cdot \Pi_{\mathbf{x}^*})   -(n-1)\Pi_z - \nabla_{z} S.
\end{equation}
In this formulation, the density function $ f $ is expressed in terms of the dual variables $(\Pi, S)$.  

Notice that if $S$ is constant, we obtain
\begin{equation}\label{density-f+}
    f =  \nabla_{\mathbf{x}} \cdot \Pi_{\mathbf{x}^*} 
    - \nabla_{\mathbf{x}^*} \cdot \Pi_{\mathbf{x}}       - \mathbf{x}^* \cdot \nabla_{\mathbf{x}^*} \Pi_z 
    + \nabla_{z}(\mathbf{x}^* \cdot \Pi_{\mathbf{x}^*})   
    - (n-1)\Pi_z,
\end{equation}
which coincides with the density function introduced in earlier studies on contact kinetic dynamics~\cite{esgu11,esgu12,EsGeGrGuPaSu}.  
Furthermore, if $\Pi_z$ vanishes identically and $\Pi_{\mathbf{x}^*}$ has no dependence on the action variable $z$, then projecting from the contact manifold $\mathbb{M}=T^*M \times \mathbb{R}$ onto the symplectic manifold $T^*M$ yields
\begin{equation}\label{density-f++}
    f = \nabla_{\mathbf{x}} \cdot \Pi_{\mathbf{x}^*} 
    - \nabla_{\mathbf{x}^*} \cdot \Pi_{\mathbf{x}},
\end{equation}
which is precisely the expression given in~\eqref{density}.

\noindent \textbf{Lie-Poisson Dynamics on the Dual Space.} Being the dual of a Lie algebra, the space $\mathfrak{g}^*_{\mathrm{ham}}$ naturally admits a Lie–Poisson bracket. 
This Lie–Poisson bracket coincides with the one given in~\eqref{LPBr} for the dual space $\mathfrak{g}^*$. 
Moreover, recalling the coadjoint action of the Lie algebra $\mathfrak{g}$ on its dual $\mathfrak{g}^*$ 
defined in~\eqref{coad-contacta}, 
we can express the coadjoint action of the Lie algebra 
$\mathfrak{g}_{\mathrm{ham}}$ on its dual space $\mathfrak{g}^*_{\mathrm{ham}}$ as
\begin{equation}\label{coad-+-+}
\begin{split}
   &ad^*: \mathfrak{g}_{\mathrm{ham}} \times \mathfrak{g}^*_{\mathrm{ham}} 
   \longrightarrow \mathfrak{g}^*_{\mathrm{ham}},\\[3pt]   
   &ad^*_{(X_H,\mathcal{R}(H))}(\Pi,S)
   =\Big(
   \mathcal{L}_{X_H}\Pi 
   - (n+1)\frac{\partial H}{\partial z}\,\Pi 
   - \mathcal{R}(H)\,dS, \\ &\qquad \qquad \qquad \qquad \qquad \qquad \qquad
   X_H(S) - (n+1)\frac{\partial H}{\partial z}\,S
   \Big),
\end{split}
\end{equation}
where we have assumed that $\mathbb{M}$ is a $2n+1$ dimensional manifold, and the divergence of the contact Hamiltonian vector field $X_H$, 
as computed in~\eqref{div-X-H-}.

We recall now the dynamics given in \eqref{LPEq-}. We assume now a Hamiltonian functional on the dual space $\mathfrak{g}^*_{\mathrm{ham}}$ as follows
\begin{equation} \label{Ham-a-}
\mathcal{H}(\Pi,S) = \int \langle \Pi, X_{H} \rangle  \mathrm{Vol}_\eta + \int  S \mathcal{R}(H)    \mathrm{Vol}_\eta .
\end{equation}
This Hamiltonian functional differs from the one defined in~\eqref{Ham-a}. 
For the Hamiltonian functional $\mathcal{H}$ in~\eqref{Ham-a-}, 
the Fr\'{e}chet derivative takes values in the Lie algebra of contact Hamiltonian vector fields 
$\mathfrak{g}_{\mathrm{ham}}$ rather than in the full Lie algebra $\mathfrak{g}$, that is,
\begin{equation}
    \frac{\delta \mathcal{H}}{\delta (\Pi,S)} \in \mathfrak{g}_{\mathrm{ham}},
    \qquad 
    \frac{\delta \mathcal{H}}{\delta \Pi} = X_H, 
    \qquad 
    \frac{\delta \mathcal{H}}{\delta S} = \mathcal{R}(H).
\end{equation} 
Hence, the equations for the kinetic (Lie-Poisson) dynamics of a bunch of particles moving through contact Hamiltonian vector fields is
\begin{equation}
\Big(\frac{d\Pi}{dt},\frac{dS}{dt}\Big)=-ad^*_{(\delta  \mathcal{H}/ \delta \Pi, \delta \mathcal{H}/\delta S)} (\Pi,S).
\end{equation} 
Referring to the coadjoint action in \eqref{coad-+-+} we have the explicit expression 
\begin{equation}\label{LPEq-+++-+-}
\begin{split}    
\frac{d\Pi}{dt}  & = -  \mathcal{L}_{X_{H}}  \Pi - \mathcal{R}(H)\big((n+1)\Pi + dS\big)   ,\\
\frac{dS}{dt}  &= - X_{H}(S) - S  (n+1) \mathcal{R}(H) S,
\end{split}
\end{equation}
where we have assumed that $\mathbb{M}$ is a $2n+1$ dimensional manifold.

Note that if $S$ is constant, then we have the Lie-Poisson equations \eqref{LPEq-+++-+-} turn out to be 
\begin{equation}\label{LPEq-+++} 
\frac{d\Pi}{dt}   = -  \mathcal{L}_{X_{H}}  \Pi - (n+1)\mathcal{R}(H) \Pi  . 
\end{equation}
This momentum dynamics is the one that exists in the literature~\cite{esgu11,esgu12,EsGeGrGuPaSu}. Moreover, if the Hamiltonian function does not depend on the action variable $z$, then we get 
 \begin{equation}\label{LPEq-++++} 
\frac{d\Pi}{dt}     = -  \mathcal{L}_{X_{H}}  \Pi  
\end{equation}
which is the momentum formulation \eqref{LP-mom-eq} of Liouville (Vlasov) equation.

 \subsection{Hamiltonian Contact Kinetic Dynamics: Density Formulation}\label{sec:Ham-den}

We consider a contact manifold $(\mathbb{M},\eta)$ endowed with the contact bracket \eqref{cont-bracket-L}. 
As discussed previously, this bracket equips the space of smooth functions on $\mathbb{M}$ with a Lie algebra structure. 
To emphasize this point, we introduce the notation
\begin{equation}
    \mathfrak{h} = C^\infty(\mathbb{M}) \simeq 
    \mathfrak{g}_{\mathrm{ham}} = \mathfrak{X}_{\mathrm{con-ham}}(\mathbb{M}).
\end{equation}
The Lie algebra structure on $\mathfrak{h}$, that is, the adjoint action, is defined by the contact bracket as 
  \begin{equation}\label{ad*oguzhan}
 ad: \mathfrak{h}\times \mathfrak{h}\longrightarrow \mathfrak{h},  \qquad ad_{H} F =\{H,F\}^{(C)}.
  \end{equation}

After fixing a contact volume form $\mathrm{Vol}_\eta$, and referring to the Lebesgue pairing, we identify the dual space of 
$\mathfrak{h} = C^\infty(\mathbb{M})$, that is  the space of densities  $
    \mathfrak{h}^* = \mathrm{Den}(\mathbb{M}),
$ with itsself the Lie algebra $\mathfrak{h} = C^\infty(\mathbb{M})$ the space of smooth function on $\mathbb{M}$. Being the dual space of a Lie algebra, $ \mathfrak{h}^*$,  inherits a Lie–Poisson structure. For two functional $\mathcal{H}$ and $\mathcal{F}$ defined on $ \mathfrak{h}^*$, the (plus) Lie-Poisson bracket \eqref{LP-bra} is 
\begin{equation}
    \{\mathcal{H},\mathcal{F}\}(f)=\int_{\mathbb{M}} f \Big\{\frac{\delta \mathcal{H}}{\delta f } ,\frac{\delta \mathcal{F}} {\delta f }\Big\}^{(C)} \mathrm{Vol}_\eta,
\end{equation}
where the bracket inside the integral is the Lie algebra operation that is the contact bracket in \eqref{cont-bracket-L}. 

Recall the isomorphism $\vartheta$ introduced in~\eqref{vartheta}, 
which maps a smooth function $H$ in $\mathfrak{h} = C^\infty(\mathbb{M})$ 
to the pair $(X_H, \mathcal{R}(H))$ in the Lie algebra 
$\mathfrak{g}_{\mathrm{ham}} = \mathfrak{X}_{\mathrm{con-ham}}(\mathbb{M})$. 
Its dual map $\vartheta^*$, defined in~\eqref{Psi-*}, is a Poisson map, 
and thus carries the Lie–Poisson structure on $\mathfrak{g}^*_{\mathrm{ham}}$ 
to the corresponding structure on $\mathfrak{h}^*$. 
Consequently, the Hamiltonian Lie–Poisson dynamics obtained in the previous subsection on $\mathfrak{g}^*_{\mathrm{ham}}$ 
can be equivalently expressed as Hamiltonian Lie–Poisson dynamics on $\mathfrak{h}^*$. 
This motivates us to study the Lie–Poisson formulation directly on $\mathfrak{h}^*$. 
Accordingly, we begin by computing the coadjoint action of the Lie algebra $\mathfrak{h}$ on its dual $\mathfrak{h}^*$.

\noindent \textbf{The Dual of Poisson Bracket Operation.} For the computation of the dynamical equation, we need the coadjoint action of the Lie algebra $\mathfrak{h}$ on its dual space $\mathfrak{h}^*$. This is done by the dualization of the adjoint action $ad$ in \eqref{ad*oguzhan}:
  \begin{equation}\label{sennur}
 ad: \mathfrak{h}\times \mathfrak{h}^*\longrightarrow \mathfrak{h}^*,  \qquad \langle ad_{H}^*f, F \rangle=\langle ad_{H} F, f \rangle . 
  \end{equation}
To get the coadjoint action, we compute  
\begin{equation}\label{cont-bra-comt}
\begin{split}
&  \langle ad_{H} F, f \rangle= -\langle ad_{F} H, f \rangle
\\ &= -\int \{F,H\}^{(C)}f \mathrm{Vol}_\eta  \\ 
& =- \int  \big( X_{H}(F)+F \mathcal{R}(H) \big)  f\mathrm{Vol}_\eta 
\\ &=-\int  f \big(\iota_{X_{H}} dF \big)      \mathrm{Vol}_\eta - \int fF \mathcal{R}(H) \mathrm{Vol}_\eta 
\\ &=-\int  f dF    \wedge  \iota_{X_{H}} \mathrm{Vol}_\eta - \int fF \mathcal{R}(H) \mathrm{Vol}_\eta 
\\ &=   \int  Fdf \wedge  \iota_{X_{H}} \mathrm{Vol}_\eta
+ \int  Ff  d\iota_{X_{H}} \mathrm{Vol}_\eta 
 - \int fF \mathcal{R}(H) \mathrm{Vol}_\eta 
 \\ &= 
  \int  F(\iota_{X_{H}} df ) \mathrm{Vol}_\eta
+ \int  Ff  \mathrm{div}(X_{H}) \mathrm{Vol}_\eta 
- \int fF \mathcal{R}(H) \mathrm{Vol}_\eta 
 \\ &= 
  \int  FX_{H}(f) \mathrm{Vol}_\eta
- (n+1) \int  Ff   \mathcal{R}(H) \mathrm{Vol}_\eta 
- \int fF \mathcal{R}(H) \mathrm{Vol}_\eta 
 \\
  &=  \int F\big(\{f,H\}^{(C)}-f\mathcal{R}(H)\big) \mathrm{Vol}_\eta   - (n+2) \int  Ff   \mathcal{R}(H) \mathrm{Vol}_\eta 
    \\
  &=  \int F \{f,H\}^{(C)} \mathrm{Vol}_\eta  -  (n+3) \int  Ff  \mathcal{R}(H) \mathrm{Vol}_\eta.
\end{split}
\end{equation}
Therefore, referring to the dualization in~\eqref{sennur} and in light of the computation~\eqref{cont-bra-comt}, 
we deduce that the coadjoint action is given by
\begin{equation} \label{coad-f}
    \operatorname{ad}^*_{H} f = \{f,H\}^{(C)} - (n+3)\, f\, \mathcal{R}(H).
\end{equation}
Notice that, as a byproduct, we obtain the following identity:
\begin{equation}
    \int_{\mathbb{M}} f\, \{H,F\}^{(C)}\, \mathrm{Vol}_\eta 
    =  
    \int_{\mathbb{M}} 
    \big( \{f,H\}^{(C)} - (n+3) f\, \mathcal{R}(H) \big) F\, \mathrm{Vol}_\eta.
\end{equation}
If the functions are independent of the action variable $z$, 
then the contact bracket reduces to the Poisson bracket, 
and the Reeb term vanishes. In that case, we have
\begin{equation}
    \int_{\mathbb{M}} f\, \{H,F\}\, \mathrm{Vol}_\eta 
    =  
    \int_{\mathbb{M}} \{f,H\}\, F\, \mathrm{Vol}_\eta,
\end{equation}
which is structurally identical to~\eqref{poissoncompo}.

As discussed in the previous subsection, the dynamics on the density level is determined through the coadjoint action. In particular, for the Hamiltonian functional 
   \begin{equation}
\mathcal{H}(f)=\int H f \mathrm{Vol}_\eta
    \end{equation}
where $H$ is the Hamiltonian function generating the particle motion on the contact manifold $\mathbb{M}$. The Fr\'{e}chet derivative $\delta \mathcal{H} / \delta f $ becomes $H$. Substituting the action in \eqref{coad-f}, the coadjoint flow is thus 
   \begin{equation}\label{LP-con-den}
   \begin{split}
\dot{f} &= - ad^*_{\delta \mathcal{H} / \delta f} f =-  ad^*_{H} f
\\&= \{H,f\}^{(C)} + (n+3)\, f\, \mathcal{R}(H)
          \end{split}
\end{equation}
 the kinetic equation in the Darboux coordinates becomes
  \begin{equation}\label{LP-con-den-exp}
  \begin{split}
\frac{\partial f}{\partial t} &=  \nabla_\mathbf{x} H \cdot \nabla _{\mathbf{x}^*}f - \nabla_{\mathbf{x}^*} H \cdot \nabla_\mathbf{x} f 
+ \big(H - \mathbf{x}^* \cdot \nabla_{\mathbf{x}^*} H\big) \nabla_z f\\ & \quad   +(\mathbf{x}^* \cdot \nabla_{\mathbf{x}^*} f) \nabla_z H + (n+2)f \nabla_z H.  
\end{split}
  \end{equation}
This evolution shows a generalization of kinetic theory to contact Hamiltonian systems, similarly as in \cite{EsGeGrGuPaSu}. Here, however, we have a more general setting, using the complete group of contact diffeomorphisms, including the conformal factors.

We remark that the Lie-Poisson dynamics \eqref{LPEq-+} in the momentum variables and the Lie-Poisson dynamics \eqref{LP-con-den}in the density formulation are related with the Poisson mapping $\Pi\mapsto f$ given in \eqref{Psi-*},

When Equation \eqref{LP-con-den-exp} is integrated over the whole contact phase space, assuming vanishing terms at all boundaries, we obtain that
\begin{equation}
  \frac{d}{dt}\int f \, \mathrm{Vol}_\eta = n \int f \frac{\partial H}{\partial z} \mathrm{Vol}_\eta,
\end{equation}
which means that the phase space volume is not conserved, as expected, due to the non-zero action of the Reeb vector field on the Hamiltonian function. Note that if $\partial H/\partial z=\mathrm{const}$, then a rescaled distribution function 
\begin{equation}
f e^{-n(\partial H/\partial z) t}
\end{equation}
turns out to be a conserved quantity. 

In summary, we have formulated a geometrical approach towards kinetic theory on contact manifolds. Since the phase-space volume is not conserved, the approach could be suitable for the study of non-equilibrium statistical mechanics, where simulations are often carried out by using non-Hamiltonian equations of motion \cite{sarman1998}, for instance, thermostats.

\section{GENERIC in Geometric Hamilton–Jacobi Theory}\label{sec.app}
In this section, we improve the geometric formulation of the GENERIC framework within contact geometry. This leads to a formalism connecting GENERIC with geometric Hamilton-Jacobi theory. 

\subsection{Contact Hamilton-Jacobi Theory} \label{sec.HJ}

We consider the projection from the contact manifold $(\mathbb{M}=T^*M \times \mathbb{R},\eta_M)$ considered to be the extended cotangent bundle to the base manifold $M$ given by 
\begin{equation}\label{pp}
\pi :T^*M  \times \mathbb{R}\longrightarrow M, \qquad (\mathbf{x},\mathbf{x}^*,z)\mapsto   \mathbf{x}. 
\end{equation}
Following \eqref{j1F}, the first jet prolongation of a real-valued smooth function $\Phi$ defined on $M$ is a section of $\pi$:
\begin{equation} \label{gamma-z}
       j^1\Phi: M  \mapsto T^* M  \times \mathbb{R},\qquad         \mathbf{x} \mapsto (\mathbf{x},\nabla_{\mathbf{x}} \Phi ,\Phi(\mathbf{x})), 
\end{equation}
image space of which is a Legendrian submanifold of $T^* M  \times \mathbb{R}$. 

We can also project the contact manifold $T^*M\times \mathbb{R}$ to the product manifold $M\times \mathbb{R}$ as follows:
\begin{equation} \label{pi1}
\pi^1:T^*M\times \mathbb{R}\longrightarrow M\times \mathbb{R}, \qquad (\mathbf{x},\mathbf{x}^*,z)\mapsto (\mathbf{x},z) .
\end{equation}

We consider a Hamiltonian function $\Psi$ on the contact manifold $T^*M \times \mathbb{R}$ and the associated contact Hamiltonian vector field $X_\Psi$.
Using $j^1\Phi$ in \eqref{gamma-z}, and the tangent mapping $T\pi$ of $\pi$ in \eqref{pp}, we project $X_\Psi$ to a vector field on the base manifold  $M$ as
\begin{equation} \label{gamma-field-C-I}
    X^\Phi_\Psi = T\pi \circ X_{\Psi} \circ j^1\Phi.
\end{equation}
The following commutative diagram then summarizes these mappings and vector fields: 
\begin{equation}\label{pic}
\xymatrix{ T^{*}M \times \mathbb{R}
\ar[dd]^{\pi} \ar[rrr]^{X_{\Psi}}&   & &T(T^{*}M \times \mathbb{R})\ar[dd]_{T\pi}\\
  &  & &\\
M   \ar@/^2pc/[uu]^{j^1\Phi}\ar[rrr]^{X_\Psi^\Phi}&  & & TM \ar@/_2pc/[uu]_{T (j^1\Phi)} }
\end{equation} 

The geometric Hamilton-Jacobi theorem reads that the following two conditions are equivalent:
\begin{enumerate}
    \item The vector field $X_\Psi^\Phi$ and $X_{\Psi}$ are $(j^1\Phi)$-related that is,
\begin{equation}\label{eq:HJ_related-1}
    X_{\Psi}\circ j^1\Phi=T (j^1\Phi)\circ X_\Psi^\Phi. 
\end{equation} 
    \item  The identity
\begin{equation}\label{eq:HJ_related-2}
    \Psi \circ j^1\Phi= 0
    \end{equation} 
    is satisfied.  
\end{enumerate}  
Note that the second condition means that the contact Hamiltonian $\Psi$ is constant on the Legendrian submanifold defined by the image space of $j^1\Phi$. 

In the Darboux coordinates $(\mathbf{x},\mathbf{x}^*,z)$, the projected  dynamics $X_\Psi^\Phi$ defined in \eqref{gamma-field-C-I} becomes
\begin{equation}\label{contact-+}
X_\Psi^\Phi=\nabla_{\mathbf{x}^*} \Psi (\mathbf{x},\nabla_{\mathbf{x}}\Phi,\Phi(\mathbf{x}))\cdot \nabla_\mathbf{x},
\end{equation}
generating the differential equation 
\begin{equation}
\dot{\mathbf{x}}=\nabla_{\mathbf{x}^*}\Psi(\mathbf{x},\nabla_{\mathbf{x}}\Phi,\Phi(\mathbf{x})),
\end{equation}
which is of the gradient type \cite{landau-ginzburg,gyarmati,pkg} with $\Psi$ as a dissipation potential. What we refer to as the contact Hamilton–Jacobi equation \eqref{eq:HJ_related-2} can be equivalently expressed as 
\begin{equation}\label{eq:HJ_related-2-}
\Psi(\mathbf{x},\Phi_\mathbf{x},\Phi(\mathbf{x}))=0.
\end{equation}  
Solving the Hamilton–Jacobi problem \eqref{eq:HJ_related-2-} consists in finding a smooth function $\Phi$ whose graph defines a Legendrian submanifold annihilating the contact one-form $\eta_M$. 

\noindent \textbf{Projected Dynamics on Graph Space.} 
We assume that we have determined a solution $\Phi$ to the Hamilton-Jacobi problem $\Psi \circ j^1\Phi= 0$. Now we determine an intermediate level to the commutative diagram presented in \eqref{pic} as follows. The graph of $\Phi$ is a submanifold of the product manifold $M\times \mathbb{R}$:
 \begin{equation}
\mathrm{gr}(\Phi)=\Big\{(\mathbf{x},\Phi(\mathbf{x}))\in M\times \mathbb{R} ~ : ~ \Phi\in C^\infty (M )\Big\}. 
 \end{equation}
Referring to the projection $\pi^1$ defined in \eqref{pi1}, from $T^*M \times \mathbb{R}$ to the product manifold $M \times \mathbb{R}$, we draw the following picture:
\begin{equation}\label{geomdiagram2}
\xymatrix{ T^*M \times \mathbb{R} \ar[dr]^{\pi^1}
\ar[ddd]^{\pi} \ar[rr]^{X_{\Psi}}&    &T(T^{*}M\times \mathbb{R})\ar[ddd]^{T\pi } |<<<<<<<<<<<<{\hole} \ar[dr]^{T\pi^1}\\
  & \mathrm{gr} (\Phi)  \subset M \times \mathbb{R} \ar@/^1pc/[ul]^{\sigma}\ar[ddl]^{\tau}\ar[rr]^{\hat{X}_\Psi^\Phi \qquad \qquad}    & &T \mathrm{gr} (\Phi) \subset T(M\times \mathbb{R}) \\\\
M \ar@/^1pc/[uuu]^{j^1\Phi}\ar@/^1pc/[uur]^{\bar{\Phi}} \ar[rr]^{X_\Psi^\Phi} &  &  T M  \ar[uur]_{T\bar{\Phi}}}
\end{equation}
In the commutative diagram, we introduce a new dynamics $\hat{X}_\Psi^\Phi$ defined on the graph of $\Phi$ as 
\begin{equation}\label{Gen-v-f-}
    \hat{X}_\Psi^\Phi:  \mathrm{gr} (\Phi) \longrightarrow T \mathrm{gr} (\Phi),\qquad (\mathbf{x},\Phi(\mathbf{x}))\mapsto T\pi^1 \circ X_\Psi \circ \sigma  (\mathbf{x},\Phi(\mathbf{x}))
\end{equation}
where $\sigma$ is the mapping
\begin{equation}
\sigma: \mathrm{gr} (\Phi) \longrightarrow  T^*M \times \mathbb{R},\qquad (\mathbf{x},\Phi(\mathbf{x})) \mapsto (\mathbf{x},\nabla_{\mathbf{x}} \Phi,\Phi(\mathbf{x}))
\end{equation}
satisfying that the composition $\pi^1\circ \sigma$ as the identity mapping. If $\Phi$ is a solution of the contact Hamilton-Jacobi problem, then $\hat{X}_\Psi^\Phi$ and $X_\Psi$ are $\sigma$-related, that is
\begin{equation}
X_\Psi \circ \sigma = T\sigma\circ \hat{X}_\Psi^\Phi. 
\end{equation}

In terms of the Darboux coordinates $(\mathbf{x},\mathbf{x}^*,z)$ on the contact manifold $T^*M \times \mathbb{R}$,  we have the dynamics $\hat{X}_\Psi^\Phi$ on the graph space as 
\begin{equation}\label{equ3}
\begin{split}
\dot{\mathbf{x}} =&~\nabla_{\mathbf{x}^*}\Psi \big\vert _{j^1\Phi}  \\
\dot{\Phi} =& ~\mathbf{x}^* \cdot \nabla_{\mathbf{x}^*}\Psi \big\vert _{j^1\Phi},
\end{split}
\end{equation}
which is simpler than the earlier geometric formulation of gradient dynamics \cite{esgrpa22a}, where also the conjugate variables had their own evolution.

\subsection{GENERIC within Geometric Hamilton-Jacobi Picture}\label{subsecBW}


We list its essential properties of the Boltzmann-Waldmann dissipation potential $\Xi=\Xi(\mathbf{x},\mathbf{x}^*)$ defined on the cotangent bundle $T^*M$, see \cite{waldmann-transport}. They are about the local behavior of the dissipation potential around the zero sections of the cotangent bundle. More concretely, we ask that: 
\begin{itemize}
    \item $\Xi$ vanishes on the zero sections of the cotangent fibration.
    \item $\Xi$ attains its minimum on the zero sections of the cotangent fibration.
    \item $\Xi$ is convex in a small neighborhood of the zero sections. 
\end{itemize}

We have assumed Boltzmann-Waldmann dissipation potential $\Xi$ as a smooth function on the cotangent bundle $T^*M$. It is possible to pullback the potential by means of the fibration $\pi^1$ in \eqref{pi1}. This reads the dissipation potential $(\pi^1)
^*\Xi=\Xi \circ \pi^1$ as a smooth function on the contact manifold $T^*M\times \mathbb{R}$. Not to cause notation inflation, we shall denote both $\Xi$ on $T^*M$ and $(\pi^1)
^*\Xi$ on $T^*M\times \mathbb{R}$ with the same symbol $\Xi$. In the contact case, the conditions listed above as follows: 
\begin{itemize}
    \item $\Xi$ vanishes on the zero sections of $\pi^1$ in \eqref{pi1}.
    \item $\Xi$ attains its minimum on the zero sections of $\pi^1$.
    \item $\Xi$ is convex in a small neighbourhood of the zero sections of $\pi^1$. 
\end{itemize}

\noindent \textbf{GENERIC as Projected Dynamics.}
We are now ready to define a contact Hamiltonian function referring to the Boltzmann-Waldmann dissipation potential $\Xi$ and a smooth function $\Phi$ as follows 
\begin{equation}\label{Psi}
\Psi (\mathbf{x},\mathbf{x}^*,z)= \Xi \big\vert_{j^1\Phi} - \Xi + \bar{\Phi}  = \Xi(\mathbf{x},\nabla_{\mathbf{x}}\Phi(\mathbf{x}),\Phi(\mathbf{x}))- \Xi(\mathbf{x},\mathbf{x}^*,z) +   \mathbf{x}^* \cdot \mathbb{L} \nabla_{\mathbf{x}}\Phi 
\end{equation}
where $\bar{\Phi}$ is the contact lift in \eqref{E-T*M-}. Evidently, a solution of the contact Hamilton-Jacobi problem for $\Psi $ in \eqref{Psi} is $\Phi$, that is
$\Psi \circ j^1\Phi =0$.

Being a solution of the contact Hamilton-Jacobi problem, the projected dynamics $X_\Psi^\Phi$ in \eqref{gamma-field-C-I} on the base manifold $M$, in Darboux coordinate, reads the GENERIC equation in form:
\begin{equation}\label{generic-pre}
\begin{split}
\dot{\mathbf{x}}&=\frac{1}{E^*}\mathbb{L}\nabla_{\mathbf{x}}\Phi-\nabla_{\mathbf{x}^*}\Xi ~ \Big \vert_{\mathbf{x}^*=\nabla_{\mathbf{x}}\Phi} . 
\end{split}
\end{equation}

Moreover, we can define the projected dynamics $\hat{X}_\Psi^\Phi$ in \eqref{Gen-v-f-} on the graph space $\mathrm{gr}(\Phi)$. In terms of the Darboux coordinates, the projected dynamics \eqref{equ3} gives the GENERIC equation in form:
\begin{equation}\label{generic}
\begin{split}
\dot{\mathbf{x}}&=\frac{1}{E^*}\mathbb{L}\nabla_{\mathbf{x}}\Phi-\nabla_{\mathbf{x}^*}\Xi ~\Big \vert_{\mathbf{x}^*=\nabla_{\mathbf{x}}\Phi} \\
\dot{\Phi}&=-\mathbf{x}^* \cdot \nabla_{\mathbf{x}^*}\Xi~\Big \vert_{\mathbf{x}^*=\nabla_{\mathbf{x}}\Phi}.
\end{split}
\end{equation}
By the symbol $\Phi$, we denote the thermodynamic potential
\begin{equation}\label{equ50}
\Phi(\mathbf{x},\mathbf{x}^*)=-S(\mathbf{x})+E^*E(\mathbf{x})+N^*N(\mathbf{x})
\end{equation}
where $S=S(\mathbf{x})$ is  the Boltzmann entropy, $N=N(\mathbf{x})$ denotes the number of moles, $E(\mathbf{x})$ is the energy. The scalars are defined to be  $E^*=1/T$ and $N^*=- \mu/T$ where $T$ is the temperature and $\mu$ chemical potential. 
Under these choices, Equation \eqref{generic} becomes the Boltzmann kinetic equation. 

\subsection{Extension of GENERIC with Microturbulence} \label{sec:micc}
In the GENERIC dynamics, thermodynamic forces enter the gradient part of the vector field in the dissipation potential. Our goal now is to pull back the GENERIC dynamics towards preceding stages in which  microscopic irregularities and instabilities  have already emerged, but they  did not yet  create thermodynamic forces driving the mesoscopic dissipation. 
The microscopic instabilities and irregularities (microturbulence) appear on the microscopic scale that is inaccessible  on the mesoscopic level as follows. 

\noindent \textbf{Particle Motion.} 
Consider two manifolds, an $m$-dimensional manifold $M$ and an $n$-dimensional manifold $N$, with local coordinates $\mathbf{x}=(x^a)$ assuming $a$ runs from $1$ to $m$, and $\boldsymbol{\xi}=(\xi^u)$ assuming $u$ runs from $1$ to $m$, respectively. In this section, we shall be using both vectorial/matrix and index notation together to be more clear about the tensorial fields that we shall introduce. We determine the product manifold 
\begin{equation}\label{adiyok}
     \widehat{M}= M \times N
\end{equation}
with coordinates $(\mathbf{x},\boldsymbol{\xi})=
(x^a,\xi^u)$. In this case, two-tuples $(\mathbf{x},\boldsymbol{\xi})$ serve as state variables, where the newly adopted $\boldsymbol{\xi}$ describes microturbulence.

We determine a bivector field on the product manifold $\widehat{M}$ in matrix notation as follows:
\begin{equation}\label{matrix}
 \widehat{\mathbb{L}} = \begin{pmatrix}
 \mathbb{L} & \mathbb{K} \\
 -\mathbb{K}^{t} & 0
 \end{pmatrix},
\end{equation}
where $\mathbb{K}^{t}$ denotes the transpose of the matrix $\mathbb{K}$. Notice that in \eqref{matrix} we require $\mathbb{L}^{(\boldsymbol{\xi})} = 0$ where $\mathbb{L}^{(\boldsymbol{\xi})}$ is the bivector expressing the kinematics of $\boldsymbol{\xi}$. This requirement follows from our interpretation of $\boldsymbol{\xi}$ as a quantity characterizing emerging microscopic irregularities in the time evolution of $\mathbf{x}$. Such quantities exist and evolve only due to the time evolution of $\mathbf{x}$.

In terms of indices, this bivector field \eqref{matrix} can be written as
\begin{equation}\label{hatL}
 \widehat{\mathbb{L}}(\mathbf{x},\boldsymbol{\xi}) = \mathbb{L}^{ij} \, \partial_{x^i} \wedge \partial_{x^j} + \mathbb{K}^{iu} \, \partial_{x^i} \wedge \partial_{\xi^u}.
\end{equation}
Hence, $\mathbb{L} = [\mathbb{L}^{ij}]$ is a skew-symmetric $m \times m$ matrix, while $\mathbb{K} = [\mathbb{K}^{iu}]$ is an $m \times n$ matrix. 
This bivector field $\widehat{\mathbb{L}}$ in \eqref{hatL} determines a bracket on the space $C^\infty(\widehat{M})$ of smooth functions. The bracket satisfies the Leibniz identity, but not necessarily the Jacobi identity. Therefore, in its present form, it defines only an almost Poisson structure.

Assume a Hamiltonian function $\Phi=\Phi(\mathbf{x},\boldsymbol{\xi},E^*,N^*)$ defined on the almost Poisson manifold $(\widehat{M},\widehat{\mathbb{L}})$ product depending also on auxiliary variables $E^*$ and $N^*$ defined in the previous subsection. Then the almost Poisson dynamics generated by the Hamiltonian function $\Phi$ is computed to be  
\begin{equation}\label{gen-1}
  \dot{x}^i= \mathbb{L}^{ij} \partial_{x^j} \Phi + \mathbb{K}^{iu} \partial_ {\xi^u}\Phi, \quad  \dot{\xi}^u= \mathbb{K}^{ui} \partial_{x^i}  \Phi.
\end{equation}
In terms of the matrix notation \eqref{matrix}, we write the dynamics in \eqref{gen-1} as
\begin{equation}\label{gen-1+}
  \dot{\mathbf{x}}= \mathbb{L}  \nabla_{\mathbf{x}}\Phi + \mathbb{K} \nabla_{\boldsymbol{\xi}}\Phi, \quad  \dot{\boldsymbol{\xi}}= - \mathbb{K}^{t} \nabla_{\mathbf{x}}\Phi.
\end{equation}

\noindent \textbf{Contact Manifold.}
We start with the product manifold $\widehat{M}=M\times N$ with local coordinates $(\mathbf{x},\boldsymbol{\xi})=
(x^a,\xi^u)$. Then determine the extended cotangent bundle as 
\begin{equation}
    \widehat{\mathbb{M}}=T^*M \times T^*N \times \mathbb{R}
\end{equation}
which is a contact manifold with contact one-form $\eta_{M\times N}$. In the Darboux' coordinates 
\begin{equation}
(\mathbf{x},\mathbf{x}^*,\boldsymbol{\xi},\boldsymbol{\xi}^*,z)= 
(x^i,x^*_i,\xi^u,\xi^*_u,z) 
\end{equation}
of the contact manifold $ \widehat{\mathbb{M}}$, the contact one-form $\eta_{M\times N}$ is 
\begin{equation}
    \eta_{M\times N}=dz - \boldsymbol{\xi}^*\cdot d \boldsymbol{\xi} - \mathbf{x}^*\cdot d \mathbf{x} , \qquad \text{or equivalently} \qquad   \eta_{M\times N}= dz - \xi^*_u d\xi^u- x^*_i dx^i.
\end{equation}
For a given Hamiltonian function $\widehat{\Psi}$ defined on the contact manifold $T^*M \times T^*N \times \mathbb{R}$, we determine the corresponding contact Hamiltonian vector field $X_{\widehat{\Psi}}$. Hence, the contact Hamiltonian dynamics is computed as 
\begin{equation}\label{equ300}
\begin{split}
\dot{\mathbf{x}}&=\nabla_{\mathbf{x}^*}\widehat{\Psi}  \\
\dot{\mathbf{x}}^*&=-\nabla_{\mathbf{x}}\widehat{\Psi}-\mathbf{x}^* \nabla_z \widehat{\Psi} \\
\dot{\boldsymbol{\xi}}&=\nabla_{\boldsymbol{\xi}^*} \widehat{\Psi}  \\
\dot{\boldsymbol{\xi}}^*&=-\nabla_{\boldsymbol{\xi}} \widehat{\Psi} - \boldsymbol{\xi}^* \nabla_z \widehat{\Psi}  \\
\dot{z}&= \mathbf{x}^*\cdot \nabla_{\mathbf{x}^*}\widehat{\Psi} +\boldsymbol{\xi}^* \cdot \nabla_{\boldsymbol{\xi}^*} \widehat{\Psi}-\widehat{\Psi}. 
\end{split}
\end{equation}
Note that for the same reason as manifold (\ref{subM}) is a Legendre and invariant manifold in the time evolution governed by (\ref{equ3}), the submanifold 
\begin{equation}\label{subhatM}
\mathrm{im}(j^1 \widehat{\Phi}) 
 =\{(\mathbf{x},\boldsymbol{\xi}, x^*,\xi^*,z)\in \widehat{\mathbb{M}} : \mathbf{x}^*=\nabla_{\mathbf{x}}\widehat{\Phi},\boldsymbol{\xi}^*=\nabla_{\boldsymbol{\xi}}\widehat{\Phi}, z=\widehat{\Phi}\}
\end{equation}
is a Legendre and invariant manifold in the time evolution governed by (\ref{equ300}).

\noindent \textbf{Contact Lift.}
Let us now construct a specific Hamiltonian function $\widehat{\Psi}$ on $T^*M \times T^*N \times \mathbb{R}$. First, recalling the discussion in Subsection \ref{lpd}, we lift the dynamics \eqref{gen-1+} to the level of cotangent bundle $T^*(M\times N)=T^*M \times T^*N$, then we push-forward this to  $T^*M \times T^*N \times \mathbb{R}$. This gives the following Hamiltonian function
\begin{equation}\label{ser}
   \widehat{E}(x^i,x^*_i,\xi^i,\xi^*_i,z) =  x_i^* \mathbb{L}^{ij} \Phi_{x^j} +  x_i^*\mathbb{K}^{iu} \Phi_{\xi^u}+   \xi^*_u \mathbb{K}^{ui} \Phi_{x^i}.
\end{equation}
In terms of the matrix notation, we have 
\begin{equation}
\widehat{E}(\mathbf{x},\mathbf{x}^*,\boldsymbol{\xi},\boldsymbol{\xi}^*,z)= \mathbf{x}^* \cdot  \mathbb{L} \nabla_{\mathbf{x}}\Phi + \mathbf{x}^* \cdot  \mathbb{K} \nabla_{\boldsymbol{\xi}}\Phi - \boldsymbol{\xi}^*\cdot \mathbb{K} ^{t} \nabla_{\mathbf{x}}\Phi. 
\end{equation}
The contact Hamiltonian vector field (the pushforward of the complete cotangent lift of the dynamics) generated by $\widehat{E}$ is then
\begin{equation}\label{contact+}
\iota_{X_{\widehat{E}}}\eta_{M\times N} =-\widehat{E},\qquad \iota_{X_{\widehat{E}}}d\eta =d\widehat{E}-\mathcal{R}(\widehat{E}) \eta_{M\times N}.   
\end{equation} 

Next, we determine a  dissipation potential $\widehat{\Xi}$ on the  contact manifold $T^*M \times T^*N \times \mathbb{R}$ in terms of the following dependencies: 
\begin{equation}\label{hatXi}
\widehat{\Xi}=\widehat{\Xi}(\mathbf{x},\boldsymbol{\xi},\boldsymbol{\xi}^*).
\end{equation}
We remark that, $\widehat{\Xi}$ depends only on the base variables $\mathbf{x}$ and $\boldsymbol{\xi}$, and the momenta $\boldsymbol{\xi}^*$. We assume also that being a dissipation potential,  $\widehat{\Xi}$ satisfies the requirements listed in Subsection \ref{subsecBW}. The physical interpretation of $\xi$ also explains the requirement (\ref{hatXi}). 

The dissipation potential generates dissipation. The role of the dissipation is to sweep away microscopic irregularities in solutions of $x$ so that an overall pattern can emerge. Since $\xi$ has been chosen to characterize the microscopic instabilities, it is $\xi$ and not $x$ that explicitly dissipates. Only the coupling of the time evolution of $\xi$ to the time evolution of $x$, that appears in $\widehat{L}$,  gives rise to  the thermodynamic forces $X$ that drive the dissipation of $x$.

To sum up, we define a contact Hamiltonian function  
\begin{equation}\label{equ500} 
\begin{split}
&\widehat{\Psi}(\mathbf{x},\mathbf{x}^*,\boldsymbol{\xi},\boldsymbol{\xi}^*,z) = \widehat{\Xi}(\mathbf{x},\boldsymbol{\xi},\boldsymbol{\xi}^*)\big\vert_{j^1\Phi} -  \widehat{\Xi} (\mathbf{x},\boldsymbol{\xi},\boldsymbol{\xi}^*)+ \widehat{E}(\mathbf{x},\mathbf{x}^*,\boldsymbol{\xi},\boldsymbol{\xi}^*,z) \\
&\qquad =\widehat{\Xi}(\mathbf{x},\boldsymbol{\xi},\Phi_\xi)- \widehat{\Xi}(\mathbf{x},\boldsymbol{\xi},\boldsymbol{\xi}^*) +  \mathbf{x}^* \cdot  \mathbb{L} \nabla_{\mathbf{x}}\Phi + \mathbf{x}^* \cdot  \mathbb{K} \nabla_{\boldsymbol{\xi}}\Phi - \boldsymbol{\xi}^*\cdot \mathbb{K} ^{\dagger} \nabla_{\mathbf{x}}\Phi
\end{split}
\end{equation}
where $\widehat{E} $ is the function defined in \eqref{ser}. We determine the contact Hamiltonian vector field $X_{\widehat{\Psi}}$. Notice that $\widehat{\Psi}$ does not depend on the action variable $z$.

\noindent \textbf{Geometry of the Passage from a Mesoscopic Level to the Level of Equilibrium Thermodynamics}. In this illustration of the Extended GENERIC we do not look at  microscopic origins of dissipation but at the geometrical interpretation of the roles that the mesoscopic thermodynamic potential $\Phi=\Phi(\mathbf{x},E^*,N^*)$ plays in the passage \textit{Mesoscopic level} $\rightarrow$ \textit{Equilibrium level} discussed  in Section \ref{subsecBW}.

We already know from the GENERIC equation \eqref{generic} that $\Phi(\mathbf{x},E^*,N^*)$ generates the time evolution making the passage. Another role that $\Phi(\mathbf{x},E^*,N^*)$ plays is in determining  the manifold
\begin{equation}\label{im1}
\mathrm{im}(j^1\Phi)=\{(\mathbf{x},\mathbf{x}^*,z)\in M`\vert\, \mathbf{x}^*=\nabla_{\mathbf{x}}\Phi,z=\Phi(\mathbf{x})\}
\end{equation}
on which the passage takes place. This result follows from lifting the GENERIC formulation \eqref{generic} of the passage to the contact GENERIC formulation (with the contact Hamiltonian $\Psi$ given in \eqref{Psi}). There are however still two other roles that   $\Phi(\mathbf{x},E^*,N^*)$ plays: it determines the fundamental thermodynamic relation (will be given in \eqref{Mesofr}) on  \textit{Mesoscopic level}, and the fundamental thermodynamic relation (will be given in  \eqref{Eqfr}) on the \textit{Equilibrium level} that is the final outcome of the passage.
Can the manifold (\ref{im1}) be extended so that  these two roles are also displayed in it?

Before answering this question we recall the essential ingredients of the mathematical formulation of the passage \textit{Mesoscopic level} $\rightarrow$ \textit{Equilibrium level}. \textit{Mesoscopic level} is represented by the state space $M$, $\mathbf{x} \in M$, and by the bivector $\mathbb{L}$ expressing  Hamiltonian kinematics of $\mathbf{x}$. The passage \textit{Mesoscopic level} $\rightarrow$ \textit{Equilibrium level} made by following solutions to  \eqref{generic-pre} is represented by
\begin{enumerate}
    \item The fundamental thermodynamic relation on  \textit{Mesoscopic level}
\begin{equation}\label{Mesofr}
S=S(\mathbf{x}); \qquad E=E(\mathbf{x}); \qquad N=N(\mathbf{x}).
\end{equation}
\item The dissipation potential $\Xi$.
\end{enumerate}
In addition, we require that $\mathbb{L}$ and $\Xi$ are degenerate in the following sense: 
\begin{equation}
\mathbb{L}\nabla_\mathbf{x} S=0, \qquad \mathbb{L}\nabla_\mathbf{x} N=0,
\end{equation}
the dissipation potential $\Xi$ depends on $\mathbf{x}^*$ only through its dependence on $K\mathbf{x}^*$, where $K$ is a linear operator satisfying 
\begin{equation}
K\nabla_{\mathbf{x}^*}E=0, \qquad K\nabla_{\mathbf{x}^*}N=0.
\end{equation}
\textit{Equilibrium level}, reached  by following solutions to \eqref{generic-pre}, is represented by the state space $M_{eq}$, $(E,N)$ in $M_{eq}$ and by the fundamental thermodynamic relation on the \textit{Equilibrium level}
\begin{equation}\label{Eqfr}
S^*=S^*(E^*,N^*)=\Phi(\mathbf{x})|_{\mathbf{x}=\mathbf{x}_{eq}(E^*,N^*)}
\end{equation}
where $\mathbf{x}_{eq}(E^*,N^*)$ is a solution to $\nabla_\mathbf{x}\Phi=0$; the thermodynamic potential $\Phi(\mathbf{x})$ is given in \eqref{equ50}.

Our objective is to display (\ref{Mesofr}) and (\ref{Eqfr}) in a manifold  (an extension of the  manifold (\ref{im1})) on which the passage \textit{Mesoscopic level} $\rightarrow$ \textit{Equilibrium level} takes place.

We make a contact lift of the extension \eqref{adiyok} with $N=M^*_{eq}; (E^*,N^*)\in M^*_{eq}, \mathbb{K}\equiv 0$  and with $\widehat{\Phi}=\Phi, \widehat{\Psi}=\Psi$, where $\Phi(\mathbf{x},E^*,N^*)$ is given in \eqref{equ50} and $\Psi(\mathbf{x},E^*,N^*)$ in \eqref{Psi}. The time evolution on the invariant manifold
\begin{eqnarray}\label{im2}
\mathrm{im}(j^1\widehat{\Phi})&=&\Big\{(\mathbf{x},\mathbf{x}^*,(E,N),(E^*,N^*))\in\widehat{M}: \mathbf{x}^*=\nabla_\mathbf{x}\widehat{\Phi},E=\nabla_{E^*}\widehat{\Phi}=E(\mathbf{x}), \nonumber \\&&N=\nabla_{N^*}\widehat{\Phi}=N(\mathbf{x}), z=-S(\mathbf{x})+E^*E(\mathbf{x})+N^*N(\mathbf{x}) \Big\}
\end{eqnarray}
is governed by the GENERIC equation \eqref{generic} and  by $\dot{E}=0,\dot{N}=0$. The manifold (\ref{im2}) displays indeed the fundamental thermodynamic relation (\ref{Mesofr}) on the  \textit{Mesoscopic level} and  the fundamental thermodynamic relation (\ref{Eqfr}) on the \textit{Equilibrium level} (on the manifold $im(j^1\hat{\Phi})|_{\mathbf{x}^*=0}$).

\noindent \textbf{Geometric Hamilton-Jacobi Theory.} The contact Hamilton-Jacobi equation \eqref{eq:HJ_related-2-} turns out to be 
\begin{equation}\label{eq:HJ_related-2-+}
\widehat{\Psi}(\mathbf{x},\boldsymbol{\xi},\nabla_{\mathbf{x}}\widehat{\Phi},\nabla_\xi \widehat{\Phi},\widehat{\Phi}(\mathbf{x},\boldsymbol{\xi}))=0, 
\end{equation}
where we are looking for a solution that is the Hamiltonian function $\widehat{\Phi}$. Accordingly, we can define projections of $X_{\widehat{\Psi}}$ to the base manifold $M\times N$, and to the graph space $\mathrm{im}(\widehat{\Phi})$, and the lifts of the these dynamics are precisely $X_{\widehat{\Psi}}$. Let us comment these one by one. 

The first jet prolongation of the Hamiltonian function $\widehat{\Phi}$ is 
\begin{equation} \label{gamma-z2}
       j^1\widehat{\Phi}: M\times N \longrightarrow T^* M \times T^*N  \times \mathbb{R},\qquad        (\mathbf{x},\boldsymbol{\xi}) \mapsto (\mathbf{x}, \nabla_{\mathbf{x}}\Phi, \boldsymbol{\xi}, \nabla_{\boldsymbol{\xi}}\Phi,\Phi(\mathbf{x},\boldsymbol{\xi})). 
\end{equation}
The image space $\mathrm{im}(\widehat{\Phi})$ is a Legendrian submanifold of the contact manifold $T^* M \times T^*N  \times \mathbb{R}$. The projection of the dynamics $X_{\widehat{\Psi}}$ on the contact manifold to the base product manifold $M\times N$ is 
\begin{equation} \label{gamma-field-C-I+}
X^{\widehat{\Phi}}_{\widehat{\Psi}}:M\times N\longrightarrow T(M\times N),\qquad   X^{\widehat{\Phi}}_{\widehat{\Psi}} = T\widehat{\pi} \circ X_{\widehat{\Psi}} \circ j^1\widehat{\Phi}.
\end{equation}
Notice that if we lift this dynamics to $T^* M \times T^*N  \times \mathbb{R}$ we arrive at $X_{\widehat{\Psi}}$, that is, they are $j^1\widehat{\Phi}$-related:
\begin{equation}
 X_{\widehat{\Psi}}   \circ j^1\widehat{\Phi} = Tj^1\widehat{\Phi}\circ X^{\widehat{\Phi}}_{\widehat{\Psi}}
\end{equation}
The projected dynamics $X^{\widehat{\Phi}}_{\widehat{\Psi}}$ in \eqref{gamma-field-C-I+} is then
\begin{equation}\label{exgeneric}
\begin{split}
\dot{\mathbf{x}}&=\mathbb{L}\widehat{\Phi}_{\mathbf{x}}+ \mathbb{K}  \widehat{\Phi}_{\boldsymbol{\xi}}  \\
\dot{\boldsymbol{\xi}}&= -\mathbb{K} ^{t} \widehat{\Phi}_{\mathbf{x}} -\widehat{\Xi}_{\boldsymbol{\xi}^*}\big \vert_{\boldsymbol{\xi}^*=\widehat{\Phi}_{\boldsymbol{\xi}}},
\end{split}
\end{equation}
which is of the GENERIC form. 

We now define a dynamics on the graph of $\widehat{\Phi}$ as follows:
\begin{equation} \label{gamma-field-C-I+-}
\hat{X}^{\widehat{\Phi}}_{\widehat{\Psi}}: \mathrm{gr}(\widehat{\Phi}) \longrightarrow T \mathrm{gr}(\widehat{\Phi}), \quad (\mathbf{x},\boldsymbol{\xi},\Phi(\mathbf{x},\boldsymbol{\xi}))\mapsto T\widehat{\pi}^1 \circ X_{\widehat{\Psi}} \circ \widehat{\sigma}  (\mathbf{x},\boldsymbol{\xi},\Phi(\mathbf{x},\boldsymbol{\xi})),
\end{equation}
where the $\widehat{\sigma} $ mapping is 
\begin{equation}
\widehat{\sigma} : \mathrm{gr} (\widehat{\Phi}) \longrightarrow  T^*M \times T^*N \times \mathbb{R},\qquad (\mathbf{x},\boldsymbol{\xi},\Phi(\mathbf{x},\boldsymbol{\xi})) \mapsto (\mathbf{x},\boldsymbol{\xi},\widehat{\Phi}_{\mathbf{x}}, \widehat{\Phi}_{\boldsymbol{\xi}}, \widehat{\Phi})
\end{equation}
satisfying that the composition $\widehat{\pi}^1\circ \widehat{\sigma}$ as the identity mapping. We remark that the contact Hamiltonian vector $X _{\widehat{\Psi}} $ and the projected vector field $\hat{X}^{\widehat{\Phi}}_{\widehat{\Psi}} $ are $\widehat{\sigma}$-related: 
\begin{equation}
    X _{\widehat{\Psi}}\circ \widehat{\sigma} =  T\widehat{\sigma}\circ \hat{X}^{\widehat{\Phi}}_{\widehat{\Psi}}.
\end{equation}
Therefore, we have the projected dynamics 
\begin{equation}\label{exgeneric.geo}
\begin{split}
\dot{\mathbf{x}}&=\mathbb{L}\widehat{\Phi}_{\mathbf{x}}+ \mathbb{K}  \widehat{\Phi}_{\boldsymbol{\xi}},  \\
\dot{\boldsymbol{\xi}}&= -\mathbb{K} ^{t} \widehat{\Phi}_{\mathbf{x}} -\widehat{\Xi}_{\boldsymbol{\xi}^*}\big \vert_{\boldsymbol{\xi}^*=\widehat{\Phi}_{\boldsymbol{\xi}}} , \\
\dot{\widehat{\Phi}}&=- \boldsymbol{\xi}^*\cdot \widehat{\Xi}_{\boldsymbol{\xi}^*} \big \vert_{\boldsymbol{\xi}^*=\widehat{\Phi}_{\boldsymbol{\xi}}}  ,
\end{split}
\end{equation}
which we may call an extended GENERIC equation. 

A diagram summarizing all the projected dynamics and discussions done so far follows: 
\begin{equation}\label{geomdiagram}
\xymatrix{ T^*M \times T^*N\times \mathbb{R} \ar[dr]^{\widehat{\pi}^1}
\ar[ddd]^{\widehat{\pi}} \ar[rr]^{X_{\widehat{\Psi}}}&    &T(T^{*}M\times T^*N\times \mathbb{R})\ar[ddd]^{T\widehat{\pi} } |<<<<<<<<<<<<{\hole} \ar[dr]^{T\widehat{\pi}^1}\\
  & \mathrm{gr} (\widehat{\Phi})  \subset M\times N \times \mathbb{R} \ar@/^1pc/[ul]^{\widehat{\sigma}}\ar[ddl]^{\tau}\ar[rr]^{\hat{X}_{\widehat{\Psi}}^{\widehat{\Phi}} \qquad \qquad}    & &T \mathrm{gr} (\widehat{\Phi}) \subset T(M\times N \times \mathbb{R}) \\\\
M \times N \ar@/^1pc/[uuu]^{j^1\widehat{\Phi}}\ar@/^1pc/[uur]^{\bar{\Phi}} \ar[rr]^{X_{\widehat{\Psi}}^{\widehat{\Phi}}} &  &  T (M\times N)  \ar[uur]_{T\bar{\Phi}}}
\end{equation}

\section{Conclusion}
In this paper, we propose a new and more generalized formulation of contact kinetic dynamics, which allows for the explicit inclusion of the dynamics of the continuum volume. For the case of infinitesimal contact transformations, the kinetic dynamics is given in \eqref{LPEq-}. For the case of Hamiltonian contact particle dynamics, the kinetic dynamics is expressed in terms of momenta and volume in \eqref{LPEq-++}, and in terms of the density function in \eqref{LP-con-den-exp}.  

We then introduce an intermediate level in the contact geometric Hamilton–Jacobi theory. As illustrated in Diagram~\ref{Gen-v-f-}, this corresponds to the graph space of the solution. Within this geometric framework, we present the GENERIC system \eqref{generic}. Referring to the abstract geometries introduced in this work, we further obtain an extended version of GENERIC \eqref{exgeneric.geo}, where an additional state variable is included to model microturbulence. We also show how this extension fits naturally within the contact geometric Hamilton–Jacobi theory, as depicted in Diagram~\ref{geomdiagram}.




In future work, we plan to extend our study of irreversible dynamics in several interrelated directions.  
Below we outline these directions in greater conceptual and technical depth.  

\begin{itemize}

\item We have observed that the dynamical equations \eqref{exgeneric.geo} of GENERIC with microturbulence can be naturally recast in the framework of Lie algebroids. Lie algebroids are abstract geometric structures that, roughly speaking, combine the geometry of the tangent bundle with that of a Lie algebra in a unified picture. Consequently, the dual space of a Lie algebroid provides a natural setting for a Poisson manifold (see, for example, \cite{Atesli25,de2005lagrangian,grabowska2006geometrical,marle2008calculus}). See \cite{ajchesgrklpa} 
for a preliminary attempt of application of Lie algebroids in the context of basic chemical kinetics.
 In particular, the dynamics arising from semi-direct product constructions fits well within this type of formulation, which we aim to explore for \eqref{exgeneric.geo}  in greater detail. We cite \cite{esvd17} for a list of semi-direct product dynamics we are mainly interested in. This may lead to address a substantially more challenging problem—namely, the dynamics on manifolds with boundaries.  
Boundaries play an essential role in open systems, energy exchange, and irreversible processes, yet their proper inclusion in geometric mechanics remains subtle.

\item We seek to understand thermodynamics within a more abstract geometric framework, formulated in terms of manifold of Legendrian submanifolds. The theory of manifolds of submanifolds (or smooth maps) is itself a highly sophisticated and technically demanding subject, see for example, \cite{Eli67,Kriegl97}, lying at the interface of differential geometry and functional analysis. 
Extending contact and algebroid geometry to this setting could provide a unifying language for describing the space of all thermodynamic states, transformations, and irreversible processes as geometric flows on such manifolds.

\item We intend to  generalize the conformal Hamiltonian kinetic motion introduced in~\cite{EsGeGrGuPaSu} by allowing the conformal factor to become a  dynamical variable rather than a fixed or externally prescribed parameter.  
In this setting, the phase-space volume and hence the entropy-related variable $S$ would become intrinsic components of the dynamics.  
This generalization may open a way to a fully geometric description of entropy production, dissipation, and non-Hamiltonian effects while preserving the underlying contact structure.

\end{itemize}

\section*{Acknowledgment}
MP was supported by Czech Science Foundation, project 23-05736S. MP is a member of the Nečas centre for mathematical modeling.

\bibliographystyle{abbrv}
\bibliography{references}

\appendix

\end{document}